\documentclass[12pt,twoside,dvips]{article}
\usepackage{amssymb}
\usepackage{pifont}
\usepackage{amsfonts}
\usepackage{amsmath}
\usepackage[spanish,english]{babel}
\usepackage{graphicx}

\begin{document}

\title{Heavy quark signals from radiative corrections to the $Z^{\prime}$ boson decay in 3-3-1 models}
\author{R. Mart\'{\i}nez$\thanks{%
e-mail: remartinezm@unal.edu.co}$ \ and F. Ochoa$\thanks{%
e-mail: faochoap@unal.edu.co}$ \and Departamento de F\'{\i}sica, Universidad
Nacional, \\
%EndAName
Bogot\'{a}-Colombia}
\maketitle

\begin{abstract}
One-loop corrections to the $Z^{\prime}$ decay width are derived and analyzed in the framework of the general form of the 3-3-1 models. We identify two important sources of corrections: oblique corrections asociated to the $Z^{\prime}$ propagator through vacum polarizations induced by virtual particle-antiparticle pairs of new heavy quarks $J$, and vertex corrections to the $Z^{\prime}q\overline{q}$ vertex through virtual exchange of new $K^{Q_{1,2}}$ gauge bosons. Fixing an specific renormalization scheme, we obtain dominant oblique corrections that exhibit a quadratic dependence on the $J$ quark mass, which are absorbed into two oblique parameters: a global parameter $\rho ^{\prime} _f$ which modify the $Z^{\prime}$ decay width, and a parameter $\kappa ^{\prime}_{f}$ that define effective $Z^{\prime}$ couplings. Numerical results in an specific 3-3-1 model gives a strong contribution of the oblique corrections from about $1.3\%$ in the $d(s)$ quark channel to $10.5\%$ in the neutrino channel, for $m_{J}=2$ TeV. The vertex corrections contribute to the oblique corrections up to $1.4\%$ for the same channel and $m_{J}$ value. For $pp$ collisions at the CERN LHC collider, we find that the corrections significantly modify the shape of the cross section distributions for $e^{+}e^{-}$ and $t\overline{t}$ final states, where the distributions including the radiative corrections increases up to $1.23$ times the tree-level distribution for the dielectron events and to $1.07$ for the top events when $m_{J}=3$ TeV.    
\end{abstract}

\section{Introduction\label{sec:introduction}}

The purpose of the next generation of experiments (LHC, ILC) will be to reveal some evidence of new physics beyond the Standard Model (SM) by detecting some signal of new matter. In particular, some extensions of the SM, new massive and neutral gauge
bosons, called $Z^{\prime }$, are predicted \cite{zprimas}. The detection of a $Z^{\prime}$
resonance has became in a matter of high priority in particle physics,
which could reveal many features about the underlying unified theory.
Indirect search for this neutral boson has been carried out at LEP,
through the mixing with the $Z$ boson \cite{LEP}. Direct $Z^{\prime}$ production
has also been searched at the Tevatron \cite{CDF,TEVA}. The discovery potential
for a $Z^{\prime}$ particle has been explored at the forthcoming Large
Hadron Collider (LHC) \cite{ATLAS} in the $M_{Z^{\prime}}\approx 1-5$ TeV range.
The search for this particle has also been explored in the planned
International Liner Collider (ILC) \cite{ILC}. If a $Z^{\prime}$ signal is detected, further analyses will be neccesary in order to determine all its properties which will allow to test the compatibility of the theoretical models with the experimental data.

On the other hand, the theoretical extensions of the SM with $Z^{\prime}$ bosons may include more unknown heavy fermions, scalar Higgs bosons, or more gauge bosons, which can also be used to constraint extensions of the SM. For example, the study of the decay modes of new heavy particles would provide additional information on the nature of the extended gauge structure. In particular, the analyses of new interactions of the $Z^{\prime}$ boson with the new fermion or Higgs content will allow to probe further details about the correct underlying theory.

There are many theoretical models which predict a $Z^{\prime}$ mass in the
TeV level, where the most popular are the $E_{6}$ motivated models \cite{zprimas,E6},
the Left-Right Symetric Model (LRM) \cite{LR}, the $Z^{\prime}$ in Little Higgs
scenario \cite{LH} and the Sequential Standard Model (SSM), which has heavier couplings than those of the SM Z boson. Searching for $%
Z^{\prime}$ in the above models has been widely studied in the literature \cite{zprimas} and
applied at LEP2, Tevatron and LHC. On the other hand, the
models with gauge symmetry $SU(3)_{c}\otimes SU(3)_{L}\otimes U(1)_{X},$
also called 3-3-1 models \cite{ten,frampton,twelve}, arise as an interesting alternative with $Z^{\prime}$
boson and many well-established motivations. First of all, from the
cancellation of chiral anomalies \cite{anomalias} and asymptotic freedom in
QCD, the 3-3-1 models can explain why there are three fermion families.
Secondly, since the third family is treated under a different
representation, the large mass difference between the heaviest quark family
and the two lighter ones may be understood \cite{third-family}. Third, the
models have a scalar content similar to the two Higgs doublet model (2HDM),
which allow to predict the quantization of electric charge and the vectorial
character of the electromagnetic interactions \cite%
{quantum-charge,vectorlike}. Also, these models contain a natural
Peccei-Quinn symmetry, necessary to solve the strong-CP problem \cite%
{PC,PC331}. Finally, the model introduces new types of matter relevant to
the next generations of colliders at the TeV energy scales, which do not
spoil the low energy limits at the electroweak scale. In particular, the model include new heavy quarks $J$ that couple to the $Z^{\prime}$ boson and new $SU(3)_L$ gauge bosons, called $K^{Q_{1,2}}$ , which can provide additional signatures of the 3-3-1 model. 

The effect of the coupling of the new quarks with the $Z^{\prime}$ boson can lead to two possible scenarios: a.) If the $Z^{\prime}$ resonance scale is bigger than the $J$ scale, then the $Z^{\prime}$ decay to $J$ quarks would be kinematically allowed. b.) If the decay to $J$ quarks is not allowed, the quantum radiative corrections through $J$ virtual loop effects to the $Z^{\prime}$ propagator is sensitive to the mass of the $J$ quarks. Some features of the first scenario have been explored in Ref. \cite{DGMO-09} in the framework of 3-3-1 models, where $Z^{\prime}$ production is searched through fermions and Higgs events as final state at LHC. In this work we will explore the second scenario, where the new heavy quarks could induce important radiative corrections to the $Z^{\prime}$ decay width, which depend on the couplings and masses of these quarks. It is to note that the new heavy quarks introduce new physics contributions at low energy, which modify some low energy parameters, as for example the $Z$-pole observables. Thus, it is also possible to study some features of the new quarks at low energy \cite{sampayo-05}. The studies of quantum corrections have been of high priority to confirm the SM predictions, as for example the precise determination of the top-quark mass \cite{bernabeu-88,bernabeu-91}. Since the tree-level phenomenology will not be a definitive test to constraint models, the small quantum corrections will also be an important matter of study in order to distinguish the correct model. Taking into account oblique and vertex corrections through heavy $J$ quarks and new heavy $K^{Q_{1,2}}$ gauge bosons, we perform a precision analysis for the 3-3-1 model, where leading one-loop contributions are considered in the $Z^{\prime}$ decay width. 

The paper has the following plan. In Sec. \ref{sec:spectrum}, we briefly review the model and the particle spectrum. In Sec. \ref{sec:neutral-currents} we introduce the relevant couplings. The detailed analysis of the oblique and vertex one-loop corrections in 3-3-1 models is discussed in Sec. \ref{sec:corrections}, including numerical results in an specific 3-3-1 model. We also perform an analysis of the effects of the radiative corrections in the CERN LHC collider. Finally in Sec. \ref{sec:conclusions}, we summarize our main conclusions.

\section{The 3-3-1 spectrum\label{sec:spectrum}}

The fermionic structure is shown in Tab. \ref{tab:tres-familias} where all
leptons transforms as $(\mathbf{3,X}_{\ell }^{L})$ and $(\mathbf{1,X}_{\ell
}^{R})$ under the $\left(SU(3)_{L},U(1)_{X}\right) $ sector, with $\mathbf{X}%
_{\ell }^{L}$ and $(\mathbf{X}_{\ell }^{R})$ the $U(1)_{X}$ generators
associated with the left- and right-handed leptons, respectively; while the
quarks transforms as $(\mathbf{3}^{\ast }\mathbf{,X}_{q_{1,2}}^{L})$, $%
(\mathbf{1,X}_{q_{2,3}}^{R})$ for two families, and $(%
\mathbf{3,X}_{q_{1}}^{L})$, $(\mathbf{1,X}_{q_{1}}^{R})$ for the first
family, each one with its $U(1)_{X}$ values for the left- and right-handed
quarks. The quantum numbers $\mathbf{X}_{\psi }$ for each representation are
given in the third column from Tab. \ref{tab:tres-familias}, where the electric
charge is defined by

\begin{equation}
Q=T_{3}+\beta T_{8}+XI,  \label{charge}
\end{equation}

\noindent with $T_{3} = 1/2$diag$(1,-1,0)$, $T_{8}=(1/2\sqrt{3})$diag$%
(1,1,-2)$ and $\beta$ a free parameter which determine diferent variations of 3-3-1 models.

\begin{table}[tbp]
\begin{center}
\begin{equation*}
\scalebox{0.8}[0.9]{\  \begin{tabular}{c||c||c} \hline\hline $Quarks$ &
$Q_{\psi }$ & $X_{\psi }$ \\ \hline\hline \begin{tabular}{c}
$q_{1L}=\left( \begin{array}{c} u_{1} \\ d_{1} \\
J_{1}\end{array}\right) _{L}:\mathbf{3}$ \\ \\
$u_{1R},\;d_{1R},\;J_{1R}:\mathbf{1}$\end{tabular} &
\begin{tabular}{c} $\left( \begin{array}{c} 2/3 \\ -1/3 \\ 1/6-\sqrt{3}\beta
/2\end{array}\right) $ \\ \\ $2/3,\;-1/3,\;1/6-\sqrt{3}\beta
/2$\end{tabular} & \begin{tabular}{c} \\ $X_{q_{1}}^{L}=1/6-\beta
/2\sqrt{3}$ \\ \\ \\
$X_{u_{1}d_{1}J_{1}}^{R}=Q_{u_{1}d_{1}J_{1}}$\end{tabular} \\
\hline\hline \begin{tabular}{c} $q_{m^{\ast }L}=\left( \begin{array}{c}
d_{m^{\ast }} \\ -u_{m^{\ast }} \\ J_{m^{\ast }}\end{array}\right)
_{L}:\mathbf{3}^{\ast }$ \\ \\ $d_{m^{\ast }R},\;u_{m^{\ast }R},\;J_{m^{\ast }R}:\mathbf{1}$\end{tabular} & \begin{tabular}{c}
$\left( \begin{array}{c} -1/3 \\ 2/3 \\ 1/6+\sqrt{3}\beta
/2\end{array}\right) $ \\ \\ $-1/3,\;2/3,\;1/6+\sqrt{3}\beta
/2$\end{tabular} & \begin{tabular}{c} \\ $X_{q_{m^{\ast }}}^{L}=1/6+\beta
/2\sqrt{3}$ \\ \\ \\ $X_{d_{m^{\ast }}u_{m^{\ast }}J_{m^{\ast
}}}^{R}=Q_{d_{m^{\ast }}u_{m^{\ast }}J_{m^{\ast }}}$\end{tabular} \\
\hline\hline $Leptons$ & $Q_{\psi }$ & $X_{\psi }$ \\ \hline\hline
\begin{tabular}{c} $\ell _{nL}=\left( \begin{array}{c} \nu _{n} \\
e_{n} \\ E_{n}\end{array}\right) _{L}:\mathbf{3}$ \\ \\ $\nu
_{nR},\;e_{nR},\;E_{nR}:\mathbf{1}$\end{tabular} &
\begin{tabular}{c} $\left( \begin{array}{c} 0 \\ -1 \\ -1/2-\sqrt{3}\beta
/2\end{array}\right) $ \\ \\ $0,\;-1,\;-1/2-\sqrt{3}\beta /2$\end{tabular} &
\begin{tabular}{c} \\ $X_{\ell _{n}}^{L}=-1/2-\beta /2\sqrt{3}$ \\ \\ \\
$X_{\nu _{n}e_{n}E_{n}}^{R}=Q_{\nu _{n}e_{n}E_{n}}$\end{tabular}
\\ \hline\hline \end{tabular}}
\end{equation*}%
\end{center}
\caption{\textsf{{\protect\small Fermionic spectrum for three families, where $m^{\ast }=2,3$ and $n=1,2,3.$}}}
\label{tab:tres-familias}
\end{table}

For the scalar sector, we introduce the triplet field $\chi$ with Vacuum Expection Value (VEV) $\langle \chi \rangle ^{T}=(0,0,\nu _{\chi})$, which provides masses to the third fermionic components. In the second transition, it is necessary to introduce two triplets $\rho$ and $\eta$ with VEV $\langle \rho \rangle ^{T}=(0,\nu _{\rho},0)$ and $\langle \eta \rangle ^{T}=(\nu _{\eta},0,0)$ in order to give masses to the quarks of up- and down-type, respectively \cite{DMO-04}.
  
In the gauge boson spectrum, it is associated with the group $SU(3)_{L}$ which transform according
to the adjoint representation and are written in the form \cite{DMO-05}

\begin{equation}
\mathbf{W}_{\mu }=W_{\mu }^{\alpha }G_{\alpha }=\frac{1}{2}\left[ 
\begin{array}{ccc}
W_{\mu }^{3}+\frac{1}{\sqrt{3}}W_{\mu }^{8} & \sqrt{2}W_{\mu }^{+} & \sqrt{2}%
K_{\mu }^{Q_{1}} \\ 
\sqrt{2}W_{\mu }^{-} & -W_{\mu }^{3}+\frac{1}{\sqrt{3}}W_{\mu }^{8} & \sqrt{2%
}K_{\mu }^{Q_{2}} \\ 
\sqrt{2}K_{\mu }^{-Q_{1}} & \sqrt{2}K_{\mu }^{-Q_{2}} & -\frac{2}{\sqrt{3}}%
W_{\mu }^{8}%
\end{array}%
\right] ,  \label{3}
\end{equation}%

whose electric charges take the general form 

\begin{equation}
Q_{W}\rightarrow \left[ 
\begin{array}{ccc}
0 & 1 & \frac{1}{2}+\frac{\sqrt{3}\beta }{2} \\ 
-1 & 0 & -\frac{1}{2}+\frac{\sqrt{3}\beta }{2} \\ 
-\frac{1}{2}-\frac{\sqrt{3}\beta }{2} & \frac{1}{2}-\frac{\sqrt{3}\beta }{2}
& 0%
\end{array}%
\right] .  \label{4}
\end{equation}%

As for the gauge field associated with $U(1)_{X},$ it is represented as$\ 
\mathbf{B}_{\mu }=B_{\mu }\mathbf{I}_{3\times 3}$ which is a singlet under $%
SU(3)_{L},$ and has no electric charge. For the charged sector, we have the following mass eigenstates \cite{DMO-05}

\begin{equation}
W_{\mu }^{\pm }=\frac{1}{\sqrt{2}}\left( W_{\mu }^{1}\mp iW_{\mu
}^{2}\right) \ ;\ K_{\mu }^{\pm Q_{1}}=\frac{1}{\sqrt{2}}\left( W_{\mu
}^{4}\mp iW_{\mu }^{5}\right) \ ;\ K_{\mu }^{\pm Q_{2}}=\frac{1}{\sqrt{2}}%
\left( W_{\mu }^{6}\mp iW_{\mu }^{7}\right) .
\end{equation}

whose eigenvalues at tree-level are

\begin{equation}
M_{W^{\pm }}^{2}=\frac{g_{L}^{2}}{4}\left( \nu _{\rho }^{2}+\nu _{\eta
}^{2}\right) ,\quad M_{K^{\pm Q_{1}}}^{2}=\frac{g_{L}^{2}}{4}\left( \nu
_{\chi }^{2}+\nu _{\eta }^{2}\right) ,\quad M_{K^{\pm Q_{2}}}^{2}=\frac{%
g_{L}^{2}}{4}\left( \nu _{\chi }^{2}+\nu _{\rho }^{2}\right)
\label{charged-mases}
\end{equation}%

For the neutral sector, we have \cite{DMO-05} 

\begin{eqnarray}
A_{\mu } &=&S_{W}W_{\mu }^{3}+C_{W}\left( \beta T_{W}W_{\mu }^{8}+\sqrt{%
1-\beta ^{2}T_{W}^{2}}B_{\mu }\right) ,  \notag \\
Z_{\mu }^{\prime } &=&-\sqrt{1-\beta ^{2}\left( T_{W}\right) ^{2}}W_{\mu
}^{8}+\beta T_{W}B_{\mu },  \notag \\
Z_{\mu } &=&C_{W}W_{\mu }^{3}-S_{W}\left( \beta T_{W}W_{\mu }^{8}+\sqrt{%
1-\beta ^{2}T_{W}^{2}}B_{\mu }\right) .
\end{eqnarray}

The corresponding eigenvalues at tree-level are 

\begin{equation}
M_{A_{\mu }}^{2}=0\ ;\ M_{Z_{\mu }}^{2}\simeq \frac{g_{L}^{2}}{4C_{W}^{2}}%
\left( \nu _{\rho }^{2}+\nu _{\eta }^{2}\right) ;\ M_{Z_{\mu }^{\prime
}}^{2}\simeq \frac{g_{X}^{2}}{3T_{W}^{2}}\nu _{\chi }^{2},
\label{masas-gauge-neutro}
\end{equation}%

where the Weinberg angle is defined as \cite{DMO-05}

\begin{equation}
T_{W}\equiv \tan \theta _{W}=\frac{g_{X}}{\sqrt{g_{L}^{2}+\left( \beta
^{2}\right) g_{X}^{2}}}\ .  \label{weinberg}
\end{equation}%

$g_{L}$ and $g_{X}$ correspond to the coupling constants of the groups $%
SU(3)_{L}$ and $U(1)_{X}$ respectively.

\section{The electroweak lagrangian\label{sec:neutral-currents}}

Using the fermionic content from table \ref{tab:tres-familias}, we obtain the
following couplings in weak eigenstates at tree level \cite{CMO-06}

\begin{eqnarray}
\mathcal{L}_{D} &=&i\overline{f}\partial \hspace{-0.2cm}/f+eQ_{f}\overline{f}%
A\hspace{-0.2cm}/f+\frac{g_{L}}{2C_{W}}\left[ \overline{f}\gamma _{\mu
}\left( g_{v}^{f}-g_{a}^{f}\gamma _{5}\right) fZ^{\mu }+\overline{f}\gamma
_{\mu }\left( \widetilde{g}_{v}^{f}-\widetilde{g}_{a}^{f}\gamma _{5}\right)
fZ^{\mu \prime }\right] \notag \\
&&+\frac{g_{L}}{\sqrt{2}} \overline{U_{L}}
W^{+}\hspace{-0.6cm}/\hspace{0.4cm} D_{L}+\frac{g_{L}}{\sqrt{2}} \overline{U_{L}}\mathbf{K}_{UJ}^{+}\mathbf{%
\hspace{-0.7cm}/}\hspace{0.5cm}J_{L}+\frac{g_{L}}{\sqrt{2}}\overline{%
D_{L}}\mathbf{K}_{DJ}^{+}\mathbf{\hspace{-0.7cm}/}\hspace{0.5cm}%
J_{L}+h.c,  \label{L-Dirac-fisico}
\end{eqnarray}

with $\mathbf{K}%
_{UJ}^{\pm }\mathbf{\hspace{-0.7cm}/}\hspace{0.5cm}=diag(K^{\pm Q_{1}}%
\hspace{-0.9cm}/\hspace{0.7cm},K^{\mp Q_{2}}\hspace{-0.9cm}/\hspace{0.7cm}%
,K^{\mp Q_{2}}\hspace{-0.9cm}/\hspace{0.7cm}),$ $\mathbf{K}_{DJ}^{\pm }%
\mathbf{\hspace{-0.7cm}/}\hspace{0.45cm}=diag(K^{\pm Q_{2}}\hspace{-0.9cm}/%
\hspace{0.5cm},-K^{\mp Q_{1}}\hspace{-0.9cm}/\hspace{0.5cm},-K^{\mp Q_{1}}%
\hspace{-0.9cm}/\hspace{0.7cm}),$ $f$ any fermion flavor, $U=(u_{1},u_{2},u_{3})$, $D=(d_{1},d_{2},d_{3})$ and $J=(J_{1},J_{2},J_{3})$ the quark components. The vector and axial vector couplings are

\begin{eqnarray}
g_{v}^{U,N} &=&\frac{1}{2}-2Q_{U,N}S_{W}^{2},\qquad \qquad
g_{a}^{U,N}=\frac{1}{2},  \notag \\
g_{v}^{D,L} &=&-\frac{1}{2}-2Q_{D,L}S_{W}^{2},\qquad \quad
\;g_{a}^{D,L}=-\frac{1}{2},  \notag \\
g_{v}^{J} &=&-2\mathbf{Q}_{J}S_{W}^{2},\qquad \qquad
\quad g_{a}^{J}=0,  \label{Z-couplings}
\end{eqnarray}

with $Q_{f}$ the electric
charges, $\mathbf{Q}_{J}=(Q_{J_{1}},Q_{J_{2}},Q_{J_{3}})$ the charges of the $J$ quarks, and $N=(\nu _{1},\nu _{2},\nu _{3})$ and $L=(e_{1},e_{2},e_{3})$ the lepton components given by Tab.
\ref{tab:tres-familias}, while the $Z^{\prime }$ couplings are defined as

\begin{eqnarray}
\widetilde{g}_{v,a}^{U} &=&%
\mbox{\small$n\left[ \frac{1}{\sqrt{3}}\left( diag\left( -1,1,1\right) +\frac{\beta
T_{W}^{2}}{\sqrt{3}}\right) \pm 2Q_{U}\beta T_{W}^{2}\right] , $}  \notag
\\
\widetilde{g}_{v,a}^{D} &=&%
\mbox{\small$n\left[ \frac{1}{\sqrt{3}}\left( diag\left( -1,1,1\right) +\frac{\beta
T_{W}^{2}}{\sqrt{3}}\right) \pm 2Q_{D}\beta T_{W}^{2}\right] ,$}  \notag
\\
\widetilde{g}_{v,a}^{J} &=&%
\mbox{\small$n\left[ \frac{2}{\sqrt{3}}\left( diag\left( 1,-1,-1\right) +\sqrt{3}\mathbf{Q}_{J}\beta
T_{W}^{2}\right) \pm 2\mathbf{Q}_{J}\beta T_{W}^{2}\right] , $} \notag
\\
\widetilde{g}_{v,a}^{N} &=&%
\mbox{\small$n\left[ -\frac{1}{\sqrt{3}}-\beta
T_{W}^{2}\right] ,$} \notag
\\
\widetilde{g}_{v,a}^{L} &=&%
\mbox{\small$n\left[ -\frac{1}{\sqrt{3}}-\beta
T_{W}^{2}\pm 2Q_{L}\beta T_{W}^{2}\right] ,$} \label{Zprima-couplings}
\end{eqnarray}

with $n=g_{X}C_{W}/2g_{L}T_{W}$. As we will explain later, we are not interested in the extra leptons $E$. For this reason, these components do not appear in the above equations. It is noted that $%
g_{v,a}^{f}$ are the same as the SM definitions and $\overset{\sim }{g}%
_{v,a}^{f}$ are $\beta $-dependent couplings of $Z_{\mu }^{\prime }$ (i.e.
model dependent).

\section{The radiative corrections in $Z^{\prime }$ decay\label{sec:corrections}}

Since the neutral lagrangian for the $Z^{\prime}$ currents have the same form as the $Z$ currents, the $Z^{\prime}$ decay width have the same form as the $Z$ width, where the vector and axial couplings $g_{v,a}$ are replaced by the new couplings $\widetilde{g}_{v,a}^{f}$ given by Eq. (\ref{Zprima-couplings}), and the $Z$ mass is changed by the $Z^{\prime}$ mass. Thus, we take the following width at tree-level   

\begin{equation}
\Gamma _{0}(Z^{\prime }\rightarrow f\overline{f})=\frac{g_{L}^{2}M_{Z^{\prime
}}N_{c}^{f}}{48\pi C_{W}^{2}}%
\mbox{\small $\sqrt{1-\mu _{f}^{\prime 2}}\left[ \left( 1+\frac{\mu _{f}^{\prime 2}}{2}\right) \left(\widetilde{g}_{v}^{f} \right) ^{2}+\left( 1-\mu _{f}^{\prime 2}\right) \left(
\widetilde{g}_{a}^{f} \right) ^{2}\right]$} %
\mbox{\footnotesize $R_{QED}R_{QCD}.$}  \label{Z'-width}
\end{equation}

\noindent where $\mu _{f}^{\prime 2}=4m_{f}^{2}/M_{Z^{\prime}}^{2}$ takes into
account kinematical corrections only important for external heavy fermions. $R_{QED,QCD}$ are global final-state corrections from photons and gluons process defined in the Appendix \ref{appendixA}, which are calculated at the $M_{Z^{\prime}}$ scale. 

On the other hand, the decay ratio can be modified by one-loop corrections sensitive to heavy particles running into the loop in both the $Z^{\prime}$ propagator and $Z^{\prime }f\overline{f}$ vertex. We will consider contributions from the exotic quarks $J$ from Tab. \ref{tab:tres-familias} and the charged bosons $K^{\pm Q_{1,2}}$ which contribute to the vertex corrections. In order to reduce the number of parameters, we will consider that the quark exotic spectrum is degenerated, and $m_{J_{1,2,3}}>M_{Z^{\prime}}$ in order to avoid additional contributions to the kinematical corrections. Thus, the kinematical factor $\mu _{f}^{\prime 2}$ takes into
account corrections only important for the top quark. We also neglect any contribution due to the exotic leptons $E_{1,2,3}$ considering that in analogy with the SM, $m_{J_{1,2,3}}>m_{E_{1,2,3}}$ so that the dominant contribution is only associated to the quarks $J_{1,2,3}$. In addition, since the masses of $K^{\pm Q_{1}}$ and $K^{\pm Q_{2}}$ are dominated by the VEV of $\chi $ as we can see in Eq. (\ref{charged-mases}) ($\nu _{\chi} >> \nu _{\eta ,\rho}$), we can consider only one $K$ mass: $M_{K}=M_{K^{\pm Q_{1}}}=M_{K^{\pm Q_{2}}}$.
Thus, from 
the couplings in Eq. (\ref{L-Dirac-fisico}), we find the one-loop corrections shown in Fig. \ref%
{fig-oblicuas}, where we can distiguish two types of corrections: the oblique corrections induced by vacuum polarization diagrams (figs (1a,b,c)), and the vertex corrections mediated by the $K_{\mu}$ gauge bosons (figs. (1d,e,f)). To calculate these corrections, we need to choose an specific renormalization scheme in order to substract all the infinite parts by properly adding divergent counterterms in the lagrangian, while the finite terms contribute to the corrections. These calculations are shown in detail in the appendix \ref{appendixA}.

\subsection{The oblique corrections}

First, we consider the diagrams (1a,b,c) which modify the $Z^{\prime}$ propagator, from where we get the renormalization of the Weinberg angle and of the $Z^{\prime}$ wave function, as described at the appendix \ref{appendixA1}. Additionally, we should take into account a renormalization of the coupling constant $g_{L}$ due to the vacuum polarization of the $K_{\mu}$ gauge boson propagators, as shown in Fig. 2. Taking a suitable parametrization we find the renormalized lagrangian given by Eq. (\ref{L-neutro-corregido-3}), from where the $Z^{\prime}$ decay width into fermions becomes

\begin{equation}
\mbox{\small $\Gamma _{Z^{\prime }\rightarrow \overline{f}f}$}=%
\mbox{\large $\frac{g_{X}^{2}M_{Z^{\prime
}}N_{c}^{f}}{192\pi T_{W}^{2}}$}\mbox{\footnotesize $\rho ^{\prime} _{f}$}%
\mbox{\footnotesize $\sqrt{1-\mu _{f}^{\prime 2}}$}%
\mbox{\small $\left[
\left( 1+\frac{\mu _{f}^{\prime 2}}{2}\right) \left( \widetilde{g}_{v,eff}^{f} \right) ^{2}+\left( 1-\mu _{f}^{\prime 2}\right) \left( \widetilde{g}_{a,eff}^{f} \right) ^{2}\right]$} %
\mbox{\footnotesize $R_{QED}R_{QCD}.$}  \label{z'-width-renorm}
\end{equation}

where the oblique corrections have a quadratic $m_J$ dependence, which is absorbed into the global factorizable parameter $\rho ^{\prime} _{f}=1+\Delta \rho_{j}$, where $\Delta \rho_{j}$ is proportional to $x_{J}=(m_{J}/M_{Z^{\prime}})^{2}$ as shown in Eq. (\ref{oblique-parameter2}), and also into the effective axial and vector couplings given by Eq. (\ref{Zprima-couplings2}), which are defined in terms of the effective Weinberg angle $\overline{T_{W}^2} = T_{W}^2(1+\epsilon \Delta \rho _{j})$, with $\epsilon$ defined in Eq. (\ref{epsilon}). We see that the width in Eq. (\ref{z'-width-renorm}) is parametrized in terms of $(g_{X}/2T_{W})^2$, while the width in Eq. (\ref{Z'-width}) is in terms of $(g_{L}/C_{W})^2$. This change of the parametrization is due to the fact that we change the parametrization of the vector and axial couplings when we define the effective couplings in Eq. (\ref{Zprima-couplings2}), which do not contain the factor $n$ that appears in Eq. (\ref{Zprima-couplings}).

\subsection{The vertex corrections}

Since the $K^{Q_{1,2}}$ gauge bosons have the same couplings to fermions as $W^{\pm },$ as we can see in the lagrangian (\ref{L-Dirac-fisico}), the vertex loops in Figs. (1d,e,f) will have a similar effect than the SM case of the $Zb\overline{b}$ vertex through a $%
t\overline{t}W^{\pm }$ loop. In the SM case, the vertex correction contribute with an additional term into the traditional oblique parameter $\rho _{f}$ due to the nonzero value of the top quark mass, where a $\Delta \rho_{t}$ parameter appears in two places for the oblique corrections: in the global term $\rho _{f}=1+\Delta \rho _{t}$ similar to the oblique parameter defined in Eq. (\ref{z'-width-renorm}) but for the $Z$ decay, and in $\kappa _{f}=1+\Delta \rho _{t}/T_{W}^2$ which define the effective angle $\overline {S_{W}^2}=\kappa S_{W}^2$ introduced in the vector coupling of the $Z_{\mu}$ boson given by Eq. (\ref{Z-couplings}). Due to the $Zb\overline{b}$ vertex correction, the above parameters contain additional leading terms: a factor $-(4/3)\Delta \rho _{t}$ in the $\rho _{b}$ parameter, and a factor $(2/3)\Delta \rho _{t}$ in the $\kappa _{b}$ parameter, i.e. $\overline{\rho_{b}} =1+\Delta \rho _{t}-(4/3)\Delta \rho _{t}$ and $\overline{\kappa _{b}} =1+\Delta \rho _{t}/T_{W}^2+(2/3)\Delta \rho _{t}$ \cite{bernabeu-91,data-particle,Bardin-97,Haywood-99}. In the $Z^{\prime}$ decay, we also have an additional contribution to the oblique parameters $\rho _{f}^{\prime} =1+\Delta \rho _{j}$ and $\kappa _{f}^{\prime} =1+\epsilon \Delta \rho _{j}$ due to the $J\overline{J}K^{Q_{1,2}}$ loop. However, from the analysis done in App. \ref{appendixA2}, we find deviations to the SM predictions. These diferences can be understood as follows:

\vspace{0.5cm}

a.) The vertex correction in the SM case is not universal due to the fact that the loop contribution is sensitive to the mass of the quarks into the loop. In particular, this correction is important only for the top quark which couple only to the $b$ quark. On the other hand, in the 3-3-1 case, we are considering the three $J$ quarks into the loop, which couples to all other quarks. Since we are considering that $m_{J_{1,2,3}}$ is the same for all the three heavy quarks, the $J$ quarks couple universally to all other quarks. Thus, the vertex correction for the $Z^{\prime}$ decay should be taken into account for all $Z^{\prime}q\overline{q}$ vertices.

\vspace{0.5cm}

b.) The oblique parameter $\Delta \rho _{t}$ in the SM is defined only for the top quark and contribute one time to the $b\overline{b}$ final state for both the oblique and the vertex corrections. In the $Z^{\prime}$ decay, the three $J$ quarks contribute with the same value for each $q\overline{q}$ final state in the oblique correction; for this reason we introduce a factor of $3$ in the oblique parameter $\Delta \rho _{j}$ in Eq. (\ref{oblique-parameter1}). However, in the vertex correction each $J$ quark contribute one time for each $q\overline{q}$ final state as we can see in figs. (1d,e,f). Thus, the additional vertex contribution proportional to $\Delta \rho _{j}$ should have a factor of $1/3$ for each final state flavor in order to remove the resummation of the three $J$ quarks in the definition of $\Delta \rho _{j}$.

\vspace{0.5cm}

c.) Finally, the $Zb\overline{b}$ vertex correction in the SM contribute with a counterterm in the lagrangian proportional to the $Zt\overline{t}$ coupling. Both, the $t$ and $b$ quarks enter into the neutral lagrangian as a $SU(2)_L$ doublet with isospin $t_{3}=(1/2)diag(1,-1)$ ($+1/2$ for $t$ and $-1/2$ for $b$). Thus, the counterterm associated to the $Zt\overline{t}$ coupling is regulated by the $1/2$ factor from the isospin. Similarly, for the $Z^{\prime}q\overline{q}$ vertex we find a counterterm proportional to the $Z^{\prime}J\overline{J}$ coupling. However, the quarks enter as $SU(3)_L$ triplets through the generator $\sqrt{3}T_{8}=(1/2)diag(1,1,-2)$ ($1/2$ for $U$ and $D$ quarks, and $-2(1/2)$ for $J$ quarks). Thus, in contrast to the SM case, the counterterm associated to the $Z^{\prime}J\overline{J}$ coupling is regulated by the $2(1/2)$ factor from $T_8$, it is 2 times the SM contribution.

\vspace{0.5cm}

Thus, as discussed in b.) and c.), the vertex correction modify the oblique parameters in the same way as the SM case but with a multiplicative factor of $2/3$, i.e:

\begin{eqnarray}
\overline{\rho_{q}^{\prime}} &=&1+\Delta \rho _{j}-(8/9)\Delta \rho _{j} \notag \\
\overline{\kappa ^{\prime}_{q}} &=&1+\epsilon \Delta \rho _{j}+(4/9)\Delta \rho _{j} \label{331-vertex}
\end{eqnarray} 

for any quark flavor $q$. The above result is obtained in Eq (\ref{L-neutro-corregido-5}) of the App. \ref{appendixA2}.

\subsection{Numerical results in models with $\protect%
\beta =-1/\protect\sqrt{3}$}

In order to determine the numerical effect of the $J$ quark in the $Z^{\prime}$ branching ratios, we will consider a particular 3-3-1 model. There are two main versions of 3-3-1 models in the literature corresponding to $\beta = -\sqrt{3}$ \cite{ten,frampton} and $\beta = -1/\sqrt{3}$ \cite{twelve}. Analyses of low energy deviations of the $Z-$pole obsevables carried out in Refs. \cite{CMO-06,MO-05} provide lowest bounds on the $Z^{\prime}$ mass of about $M_{Z^{\prime}} \approx 1.5$ TeV for models with $\beta = -1/\sqrt{3}$, and $M_{Z^{\prime}} \approx 4$ TeV for $\beta = -\sqrt{3}$. On the other hand, the $\beta = -\sqrt{3}$ models points to a nonperturbative regime close to the $M_{Z^{\prime}}$ scale predicted by this model, where a Landau-like pole arises when $S_{W}^{2}=1/4$ at the $\mu \approx 4$ TeV energy scale \cite{DMP-05}. Thus, we will consider the case with $\beta = -1/\sqrt{3}$ which is protected from these nonperturbative regimes at the $M_{Z^{\prime}}$ scale. Taking into account the Renormalization Group Equations (RGE), the running constants at the $M_{Z^{\prime}}$ scale are found to be \cite{CMO-06}

\begin{eqnarray}
\alpha _{L}^{-1}(1.5 \text{ TeV}) &=&%
30.906;%
\text{\qquad }\alpha ^{-1}(1.5\text{TeV})=%
125.993;  \notag \\
\qquad \alpha _{s}(1.5\text{ TeV})&=&%
0.0853; \text{\qquad } S_{W}^{2}(1.5\text{ TeV})=0.2453,
\label{RCC}
\end{eqnarray}%

with $\alpha _{L}=g_{L}^{2}/4\pi $, $\alpha$ the fine-structure constant and $\alpha _s$ the strong coupling constant. We also use the running top quark mass $\overline{m_{t}}(1.5$ TeV$)=150.73$ GeV to calculate the kinematical factor $\mu ^{\prime 2}_f$ when $f=t$. With the above data, and using the Eq. (\ref{z'-width-renorm}) (with only oblique correction for leptons and oblique plus vertex corrections for quarks), we can calculate the total decay widths $\Gamma _{f}$ into fermions $f\overline{f}$ as a function of $m_{J_{1,2,3}}=m_{J}$. We use a representation where $(u_{1},u_{2},u_{3})=(t,u,c)$ and $(d_{1},d_{2},d_{3})=(b,d,s)$, where the SM heavy quarks $(t,b)$ are treated under a different representation, as shown in Tab. \ref{tab:tres-familias}. In order to determine the $J$ mass effect on the widths, we first define the total width as $\Gamma _{f}=\Gamma _{0f}(1+\delta _{f})$, with $\Gamma _{0f}$ the corresponding tree-level width and $\delta _{f}$ the deviation due to oblique and vertex corrections. With the above definition, we can calculate de rate as $\delta _{f}=(\Gamma
_{f}-\Gamma _{0f})/\Gamma _{0f}$. The behaviour of $\delta_ {f}$ with $m_{J}$ is displayed in Fig. \ref{fig-branching-1}-a for each fermion final state. In general, we see a strong effect of the oblique corrections to the total $Z^{\prime}$ width, which grows when $m_{J}$ grows. This strong dependence can be understood taking into account that there are three virtual quarks $J_{1,2,3}$ contributing at the same time to each final state, which enhance the effect of the corrections. The oblique loops introduce corrections from $4\%$ to more than $30\%$ in the $\ell ^{\pm}$ final states in the range $1.5$ TeV $\le m_{J} \le 3.5$ TeV. The same effect can be seen in each neutrino channel $\nu \overline{\nu}$ ($\nu =\nu _{e,\mu ,\tau}$) but with opposite sign. This means that the oblique correction for $\ell ^{\pm}$ increases the tree-level width, while the width for neutrinos decreases with the corrections. The source of this difference comes from the definition of the vector and axial-vector couplings for leptons in Eqs. (\ref{Zprima-couplings}) and (\ref{Zprima-couplings2}) for the bare and effective couplings, respectively. It is possible to verify that for any value of $m_{J}$, we obtain that $\left| \widetilde{g}_{v,a,eff}^{\ell }\right|>\left| \widetilde{g}_{v,a}^{\ell }\right|$, while $\left| \widetilde{g}_{v,a,eff}^{\nu }\right|<\left| \widetilde{g}_{v,a}^{\nu }\right|$, which produces a different sign in $\delta _{f}$. The quark states also exhibit an appreciable dependence with $m_{J}$ but reduced by different factors depending on the flavor. This reduction in part is due to the presence of the vertex corrections as we will see below. The $t\overline{t}$ state show a bigger effect than its partners $(u,c)$, which is a direct consequence of the family dependence introduced  by the group representations of the model. It is an important distinction of the 3-3-1 models. We can see a similar effect in the $b\overline{b}$ channel, which shows a bigger deviation than the $d\overline{d}(s\overline{s})$ channel. In particular, the $d(s)$ states exhibit a large suppresion of the radiative effects, where we obtain corrections from $1\%$ to $4\%$ in the range $1.5$ TeV $\le m_{J} \le 4$ TeV. This suppresion is due to the fact that the vertex corrections introduce negative factors that reduce the oblique contributions. In the case of the $d(s)$ channels the vertex contributions exhibit the biggest effect of the quark states as we will see below, which combined with the oblique corrections, we obtain a low sensitivity to the $m_{J}$, as we can see in Fig. \ref{fig-branching-1}-a.

On the other hand, to explore the deviatons due only to the vertex corrections in the quark channels, we define the vertex rate ratio $\delta _{q-vertex}=(\Gamma
_{q}-\Gamma _{q}^{obl})/\Gamma _{q}^{obl}$, where $\Gamma _{q}^{obl}$ is the width with only oblique effects. Fig. \ref{fig-branching-1}-b displays the behaviour of $\delta _{q-vertex}$ with $m_{J}$. We see that the vertex corrections have smaller effects on the $Z^{\prime}$ width than the oblique corrections and with negative sign. Thus, the vertex corrections reduce the oblique contributions. We find that for the range $1$ TeV $\le m_{J} \le 4$ TeV, the vertex diagrams introduce corrections from $0.5\%$ to $5\%$ for the $d(s)$ channels, and from $0.5\%$ to $4\%$ for the $t$ events. The other channels exhibit intermediate values. It is to note that in addition to the effect of the $J$ quarks, the vertex corrections provide evidences of the new $K^{Q_{1,2}}$ gauge bosons. In Tab. \ref{tab:decay-width} we show the values of $\Gamma _{0f},\Gamma _{f}^{obl},\Gamma _{f}, \delta _{f}$ and $\delta _{q-vertex}$ for $m_{J}=2$ TeV for each fermion channel.

\begin{table}[tbp]
\begin{center}
\scalebox{0.82}[0.9]{\begin{tabular}{c|c|c|c|c|c}
\hline
 Final states & $
\Gamma _{0f}$(GeV) & $\Gamma _{f}^{obl}$(GeV) & $\Gamma _{f}$(GeV) & $\delta _{f}(\%)$ & $\delta _{q-vertex}(\%)$ \\ \hline
$u\overline{u}(c\overline{c})$ & 2.293 & 2.391 & 2.362 & 3.02 & -1.19 \\ \hline
$t\overline{t}$ & 3.271 & 3.517 & 3.481 & 6.42 & -1.04 \\ \hline
$d\overline{d}(s\overline{s})$ & 1.963 & 1.966 & 1.938 & -1.28 & -1.42 \\ \hline
$b\overline{b}$ & 2.974 & 3.122 & 3.086 & 3.76 & -1.17 \\ \hline
$e^{+}e^{-}(\mu^{\pm},\tau^{\pm})$ & 0.665 & 0.722 & 0.722 & 8.48 & ---  \\ \hline
$\nu _{e,\mu ,\tau} \overline{\nu }_{e,\mu ,\tau}$ & 0.344 & 0.308 & 0.308 & -10.48 & ---  \\ \hline
\end{tabular}}
\end{center}
\caption{\textsf{{\protect\small Decay widths at: tree-level, with only oblique corrections, and with oblique plus vertex corrections. We also show the correction fraction of the total contributions ($\delta _{f}$) and only vertex contributions ($\delta _{q-vertex}$). The data are calculated for $m_{J}=2$ TeV}}}
\label{tab:decay-width}
\end{table}

It is possible to isolate the $J$ mass dependence in the vertex corrections by taking appropiate branching ratios. Fig. \ref{fig-branching-2}-a shows the total branching ratio $Br_ {q\overline{q}}=\Gamma _{q\overline{q} }/\Sigma \Gamma _{f\overline{f}}$, with $\Sigma \Gamma _{f\overline{f}}$ the total decay width. We see small variations of the branching with the mass $m_{J}$ due to the fact that both the oblique and vertex corrections exhibit large suppressions by the ratio. The suppresion of the vertex part arises because it is included in all the quark channel. Thus, a more appropriated branching is the ratio between each quark width and the total lepton width, where only the oblique corrections are suppresed, while the vertex effect from each quark state is not canceled out due to the fact that the lepton events do not contain vertex contributions. Fig. \ref{fig-branching-2}-b shows the behaviour of this branching with $m_{J}$. We see that the decay to top quarks exhibit bigger values than the lepton decay values due to the large coupling of the top to the neutral currents $Z^{\prime}$, which is a particular feature of the 3-3-1 models. The vertex effect is displayed in the figure, where the branching to top quarks increases from about $1.12$ to $1.18$ in the range $1.5$ TeV $\le m_{J}\le $ $4$ TeV, while the effect is smaller for the $b,u,c$ channels, which exhibit small variations in the branching with $m_{J}$. It is to note that in contrast to the top system, the branching for the other quark channels decreases with $m_{J}$, which is a consequence of the negative contribution that the vertex corrections introduce, as we see in tha last column of Tab. \ref{tab:decay-width}.

\subsection{The radiative corrections at CERN LHC collider}
 
In order to explore the consequences of the corrections of the $Z^{\prime}$ decay in $pp$ collisions in the CERN LHC collider, we calculate the differential cross sections for $Z^{\prime}$ production as a function of the di-lepton invariant mass with and without the corrections. In the electron channel, we use the kinematical cuts reported in Refs. \cite{dittmar-04}, based on an integrated luminosity $L=100$ $fb^{-1}$ of $pp$ collisions at center-of-mass energy $\sqrt{s} =14$ TeV, where events are required to have two leptons with transverse energy $E_{T}\ge 20$ GeV within the pseudorapidity range $\left| \eta \right| \le 2.5$. Additionally, the lepton should be isolated within a cone of angular radius $\Delta R=0.5$ around the lepton \cite{dittmar-04}. For this study, we use the CALCHEP package \cite{calchep} in order to simulate $pp \rightarrow \ell \overline{\ell}$ events for $\ell =e$ with the above kinematical criteria. We also use a nonrelativistic Breit-Wigner function and CTEQ6M parton distribution functions \cite{cteq}. In Fig. \ref{fig-distr}-a we show the ratio of one-loop to tree-level invariant mass distribution for the di-lepton system as a final state, where we have chosen a central value $M_{Z^{\prime}}=1.5$ TeV, and $m_{J}=2$ TeV and $m_{J}=3$ TeV. As expected, there is a low influence of the final state corrections to the shape of the distributions for invariant masses far from the central value $1.5$ TeV. We find that the final state corrections dominate over the resonance range, where the one-loop distribution exhibit bigger values than the tree-level distribution, as we can see in Fig. \ref{fig-distr}-a. In the peak at $M_{Z^{\prime}}$ we obtain the biggest deviation, where the one-loop distribution is $1.09$ times the tree-level value for $m_{J}=2$ TeV and $1.23$ times the tree-level distribution for $m_{J}=3$ TeV. Thus, the mass of the $J$ quarks significantly change the cross section distributions, where the effects of the corrections increase appreciably when $m_{J}$ takes larger values.

On the other hand, due to the large number of top events that will be produced at the LHC (of about $80\times 10^{6}$ pair events), and the preferential coupling to top quarks that the model exhibit, it is interesting to study the channel $Z^{\prime} \rightarrow t\overline{t}$. We use the basic cuts $p_{T}(t) \ge 100$ GeV for transverse momentum within the rapidity range $\left| y(t) \right| \le 2.5$ \cite{han-04,baur-07}. A detailed discussion about the high invariant mass $t\overline{t}$ event reconstruction from different top decay modes can be found in Refs. \cite{baur-07,broo-08}. Fig. \ref{fig-distr}-b displays the distribution ratio for the top system, where we have used the same $M_{Z^{\prime}}$ and $m_{J}$ values. Although the $Z^{\prime}$ signal from top events is reduced due to the large QCD background, a small peak is identified for $M=1500$ GeV, as Fig. \ref{fig-distr}-b shows. For $m_{J}=2$ TeV, we find that the one-loop distribution is about $1.03$ times the tree-level values in the peak, while it is $1.07$ times the tree-level distribution for $m_{J}=3$ TeV.

\section{Conclusions\label{sec:conclusions}}

In this paper we have derived the one-loop corrections to the $Z^{\prime}$ decay in the 3-3-1 models, where we found two sources of corrections: oblique corrections to the $Z^{\prime}$ propagator due to virtual pair production of heavy quarks $J$, and vertex corrections induced by extra $K^{Q_{1,2}}$ gauge bosons which change the decay width into quark events. Taking into account dominant contributions, we found a quadratic dependence on the $J$ quark mass which are absorbed into effective coupling constants. Numerical results was obtained in the framework of the 3-3-1 model with $\beta=-1/\sqrt{3}$. Due to the fact that the $Z^{\prime}$ current couples strongly to top quarks, we found larger radiative contributions in the top channel than in other quark events, which is a particular feature of the 3-3-1 model. The lepton channel exhibit a strong dependence with $m_{J}$ values, where the oblique loops introduce corrections from $4\%$ to about $30\%$ in the range $1.5$ TeV $\le m_{J} \le 3.5$ TeV. On the other hand, we obtained smaller effects from the vertex diagrams, where corrections from $0.5\%$ to $5\%$ were obtained in the range $1$ TeV $\le m_{J} \le 4$ TeV for quark events. Using the kinematical cuts expected at the CERN LHC collider, we explored the $m_{J}$ effect on the differential cross section distribution due to the one-loop radiative corrections for $e^{+}e^{-}$ and $t\overline{t}$ final states. We found that the final state corrections significantly influences the shape of the distributions, as a large fraction of the events shifts from the peak region at the invariant mass value of $1500$ GeV to lower fractions for other mass values. The one-loop distribution increases up to $1.09$ and $1.23$ times the tree level $m(e^{+}e^{-})$ distribution for $m_{J}$ equal to $2$ TeV and $3$ TeV, respectively; and to $1.03$  and $1.07$ times the tree-level $m(t\overline{t})$ distribution for the same $m_{J}$ values.

The exploration of this model at LHC is interesting because although there are many models containing additional $U(1)$ symmetries with known phenomenology, there are few in which the $U(1)$ is family non-universal. Our calculation allow to probe further details about the 3-3-1 models and provide new tests to distinguish this model from other theoretical models with new gauge bosons.   

This work was supported by COLCIENCIAS.  
  
\section*{Appendix}

\appendix

\section{Renormalization scheme\label{appendixA}}

The $Z^{\prime }$ decay in Eq. (\ref{Z'-width}) contains global QED and QCD corrections through the definition of $R_{QED}=1+\delta _{QED}^{f}$ and $R_{QCD}=1+(1/2)%
\left( N_{c}^{f}-1\right) \delta _{QCD}^{f}$, where \cite{data-particle}, \cite{Bardin-97}, \cite{bernabeu-91}

\begin{eqnarray}
\delta _{QED}^{f} &=&\frac{3\alpha Q_{f}^{2}}{4\pi };  \notag \\
\delta _{QCD}^{f} &=&\frac{\alpha _{s}}{\pi }+1.405\left( \frac{\alpha _{s}}{%
\pi }\right) ^{2}-12.8\left( \frac{\alpha _{s}}{\pi }\right) ^{3}-\frac{%
\alpha \alpha _{s}Q_{f}^{2}}{4\pi ^{2}}  \label{QCD}
\end{eqnarray}%
with $\alpha $ and $\alpha _{s}$ the electromagnetic and QCD constants, respectively. The values $\alpha $ and $\alpha _{s}$ are calculated at the $M_{Z^{\prime}}$ scale.

\begin{figure}[t]
\centering \includegraphics[scale=0.9]{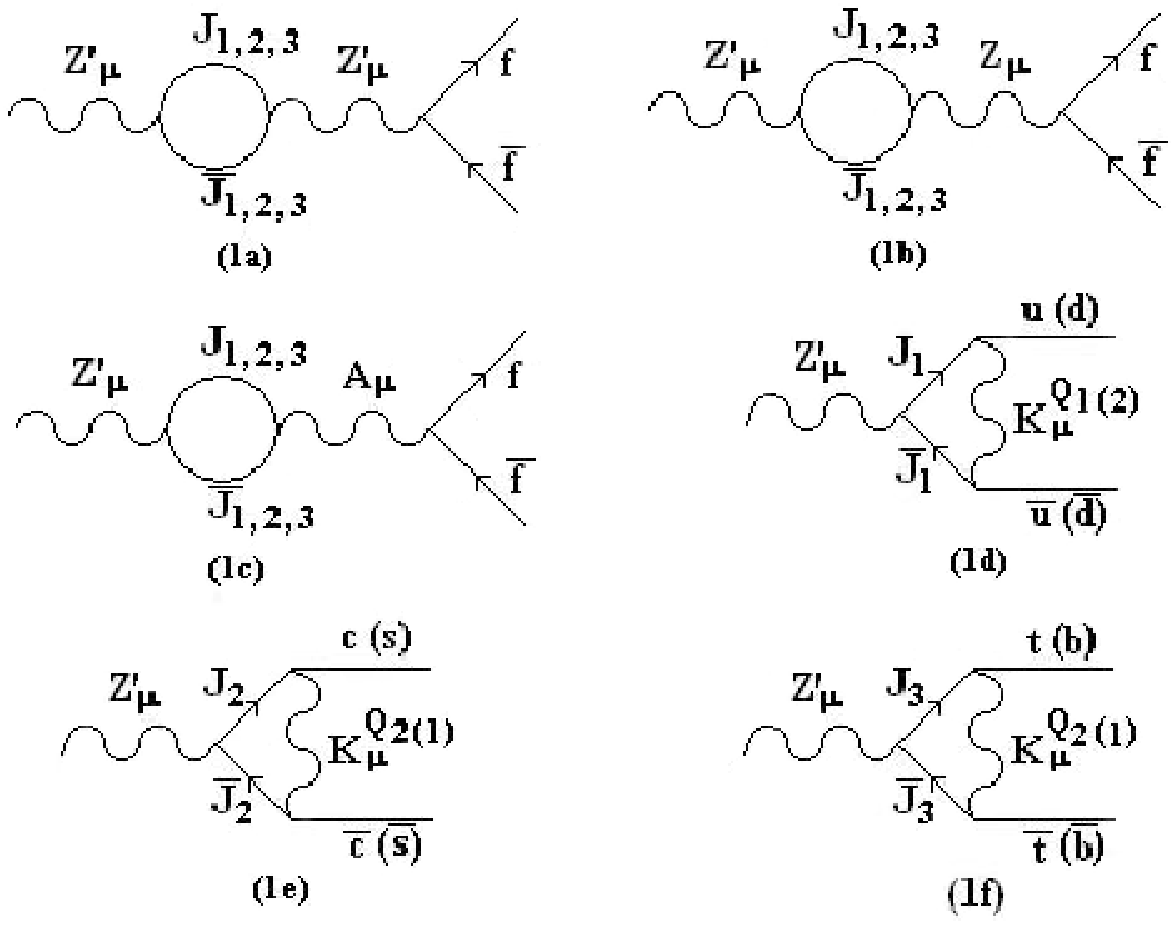}
\caption{\textsf{\ {\protect\small One loop oblique (a,b,c) and vertex (d,e,f) corrections through $J$ quarks and $K_{\mu}$ gauge bosons of the $Z^{\prime }$ decay}}}
\label{fig-oblicuas}
\end{figure}

We are also considering oblique and vertex corrections sensitive to the extra quarks masses $m_{J_{1,2,3}}$. As described in section \ref{sec:corrections}, we assume that $M_{Z^{\prime }}<m_{J_1}=m_{J_2}=m_{J_3}$. As we can see in Fig. \ref{fig-oblicuas}, we get vacuum $Z^{\prime }-Z$, $Z^{\prime }-photon$ polarizations and $Z^{\prime }$ self-energy, and vertex corrections in the $Z^{\prime }$ decay.

\subsection{Oblique corrections\label{appendixA1}} 

Let us write the following general coupling of two neutral vector bosons $T_{\mu}$ and $Y_{\mu}$ that interacts with fermions $f$
 
\begin{equation*}
\mathcal{L}_{D}^{NC}=g_{T}\overline{f}\gamma ^{\mu }\left( v^{f}-a^{f}\gamma
_{5}\right) fT_{\mu }+g_{Y}\overline{f}\gamma ^{\mu }\left( v^{\prime
f}-a^{\prime f}\gamma _{5}\right) fY_{\mu }.
\end{equation*}

Thus, the one loop self-energy with two fermiones $f\overline{f}$ running into the loop is
(in the dimensional regularization scheme) \cite{novikov-93}

\begin{eqnarray}
-i\Sigma _{TY}^{\mu \nu }(p^{2}) &=&(-1)\int \frac{d^{D}p}{(2\pi )^{D}}Tr%
\left[ \frac{g_{T}\gamma ^{\mu }\left( v^{f}-a^{f}\gamma _{5}\right) }{p%
\hspace{-0.15cm}/-m_{f}}\frac{g_{Y}\gamma ^{\nu }\left( v^{\prime
f}-a^{\prime f}\gamma _{5}\right) }{p\hspace{-0.15cm}/+k\hspace{-0.15cm}%
/-m_{f}}\right]   \notag \\
&=&-ip^{2}P^{\mu \nu }\frac{g_{T}g_{Y}N_{c}^{f}}{12\pi ^{2}}\left\{
v^{f}v^{\prime f}\left[ \Delta _{f}+(1+2x_{f})F(x_{f})-\frac{1}{3}\right] \right.  
\notag \\
&&+\left. a^{f}a^{\prime f}\left[ 3\left( 1-2x_{f}\right) \Delta
_{f}+(1-4x_{f})F(x_{f})+3+2Ln(x_{f})\right] \right\} ,  \label{V-V'-self}
\end{eqnarray}%

where $P^{\mu \nu }=g^{\mu \nu }-p^{\mu }p^{\nu }/p^{2},$ $\Delta
_{f}=2/\varepsilon -\gamma _{E}+\ln (4\pi \mu ^{2}/m_{f}^{2}),$ $\varepsilon
=(4-D)/2,$ $\gamma _{E}=0.577,$ $x_{f}=m_{f}^{2}/p^{2}$ and $F(x_{f})=2(1-\sqrt{4x_{f}-1}%
\arcsin (1/4x_{f}))$ if $4m_{f}^{2}>p^{2}.$ It is to note that the longitudinal term $p_{\mu }p_{\nu }/p^{2}$ cancel out with the corresponding Goldstone boson after symmetry breaking. We can also obtain the vacuum polarization through the equation $%
\Sigma _{TY}^{\mu \nu }=p^{2}P^{\mu \nu }\Pi _{TY}$. If we consider very heavy fermions into the loops, we can take $x_{f}\gg
1/4$, so that $F(x_{f})\rightarrow 1/(6x_{f})$. Thus, the $Z^{\prime }-Z^{\prime }$ self-energy, and the $Z^{\prime }-Z$ and $Z^{\prime }-photon$ polarizations in the propagator correction in Figs. (1a,b,c) are 

\begin{eqnarray}
\Sigma _{Z^{\prime }Z^{\prime }}(p^{2}) &=&p^{2}\Pi _{Z^{\prime
}}=p^{2}\left( \frac{g_{L}}{4\pi C_{W}}\right) ^{2}x_{J}\left\{ \left( 
\widetilde{g}_{v}^{J}\right) ^{2}\left[ \frac{\Delta _{J}}{x_{J}}+\frac{1}{6x_{J}^{2}%
}\right] \right.   \notag \\
&&-\left. \left( \widetilde{g}_{a}^{J}\right) ^{2}\left[ 6\Delta _{J}-\frac{1%
}{x_{J}}\left( 3\Delta _{J}+\frac{7}{3}\right) -\frac{1}{6x_{J}^{2}}-\frac{2}{x_{J}}\ln
(x_{J})\right] \right\} ,  \label{Z'-Z'-self}
\end{eqnarray}%

\begin{equation}
\Pi _{Z^{\prime }Z}(p^{2})=\left( \frac{g_{L}}{4\pi C_{W}}\right)
^{2}x_{J}\left\{ \widetilde{g}_{v}^{J}g_{v}^{J}\left[ \frac{\Delta _{J}}{x_{J}}+%
\frac{1}{6x_{J}^{2}}\right] \right\} ,  \label{Z'-Z-polar}
\end{equation}%

\begin{equation}
\Pi _{Z^{\prime }\gamma }(p^{2})=\frac{eg_{L}Q_{J}}{8\pi ^{2}C_{W}}x_{J}\left\{ 
\widetilde{g}_{v}^{J}\left[ \frac{\Delta _{J}}{x_{J}}+\frac{1}{6x_{J}^{2}}\right]
\right\} ,  \label{Z'-foton-polar}
\end{equation}%

with $x_{J}=m_{J}^{2}/p^{2}$ and $\widetilde{g}_{v,a}^{J}$ given by Eqs. (\ref%
{Zprima-couplings}). Since we will consider the limit $x_{J}=m_{J}^{2}/p^{2}\gg 1/4,$ which at the $Z^{\prime }$ scale becomes $m_{J}^{2}\gg M_{Z^{\prime }}^{2}/4,$ we can assume that $%
M_{Z^{\prime },K}^{2}\rightarrow 0$ in order to obtain the dominant corrections that depend of the $J$ quark masses. With these considerations, we can see from Eqs. (\ref{Z'-Z-polar}) and (\ref{Z'-foton-polar}) that $\Pi _{Z^{\prime }Z}$ and $\Pi _{Z^{\prime }\gamma }$ are negligible, while $\Pi _{Z^{\prime }}$ in (\ref{Z'-Z'-self}) contribute only to the axial component as

\begin{eqnarray}
\Pi _{Z^{\prime }} &\approx &-\left( \frac{g_{L}}{4\pi C_{W}}\right)
^{2}\left( \widetilde{g}_{a}^{J}\right) ^{2}\left[ 6x_{J}\Delta _{J}-2\ln (x_{J})%
\right].  \label{polar-leading}
\end{eqnarray}%

The above corrections modify some quantities of the theory depending on the renormalization scheme. We fix our scheme considering that the tree-level relation $M_{K0}^{2}/M_{Z^{\prime}0}^{2}=3g_{L0}^{2}T_{W0}^{2}/4g_{X0}^{2}$ which is obtained from Eqs. (\ref{charged-mases}) and (\ref{masas-gauge-neutro}), and where the subscript $0$ denotes bare quantities, takes the same form for renormalized quantities, i.e 

\begin{equation}
M_{K}^2/M_{Z^{\prime}}^{2}=3g_{L}^{2}T_{W}^{2}/4g_{X}^{2}, \label{scheme}
\end{equation}

with $g_L$, $g_X$, $T_{W}$, $M_{K}$ and $M_{Z^{\prime}}$ renormalized quantities, which are obtained as follows

\vspace{0.3cm}

\ding{172} \underline{\textbf{$Z^{\prime}$ mass and wave function renormalization}}: Since the $Z^{\prime }$ self-energy contains contributions from the 3 quarks $J_{1,2,3}$, we must introduce a multiplicative factor of 3. Thus, we obtain for the renormalized propagator

\begin{equation*}
-iG_{Z^{\prime }}^{\mu \nu }=\frac{iP^{\mu \nu }}{p^{2}-M_{Z^{\prime
}0}^{2}+3\Sigma _{Z^{\prime }Z^{\prime }}(p^{2})}\approx \frac{iP^{\mu \nu }}{%
p^{2}-M_{Z^{\prime }}^{2}}\left( 1+3\Sigma _{Z^{\prime }Z^{\prime }}^{\prime
}(M_{Z^{\prime }})\right) ^{-1},
\end{equation*}%

where in the secod equality we use the expansion of $\Sigma _{Z^{\prime }Z^{\prime }}(p^{2})$ around $M_{Z^{\prime}}^{2}$, and where the renormalized mass and wave function are defined as

\begin{eqnarray}
M_{Z^{\prime }0}^{2} &=&M_{Z^{\prime }}^{2}\left( 1+3\Pi _{Z^{\prime
}}(M_{Z^{\prime }}^{2})\right) ,  \notag \\
Z_{0}^{\prime } &=&Z^{\prime }\left( 1+3\Sigma _{Z^{\prime }Z^{\prime
}}^{\prime }(M_{Z^{\prime }}^{2})\right) ^{-1/2}=Z^{\prime }\mathcal{Z}%
_{3}^{-1/2},  \label{mass-z'-renorm}
\end{eqnarray}%

with $M_{Z^{\prime }0}^{2}$ and $Z_{0}^{\prime }$ the corresponding bare quantities and $\mathcal{Z}_{3}$ the renormalization constant. The expression for $\Sigma _{Z^{\prime }Z^{\prime }}^{\prime }$ can be obtained by derivating the self-energy from Eq. (\ref{Z'-Z'-self})

\begin{eqnarray}
\Sigma _{Z^{\prime }Z^{\prime }}^{\prime }(M_{Z^{\prime }}^{2}) &=&\left. 
\frac{d\Sigma _{Z^{\prime }Z^{\prime }}}{dp^{2}}\right\vert
_{p^{2}=M_{Z^{\prime }}^{2}}=\left( \frac{g_{L}}{4\pi C_{W}}\right)
^{2}x_{J}\left\{ \left( \widetilde{g}_{v}^{J}\right) ^{2}\left[ \frac{\Delta _{J}%
}{x_{J}}+\frac{1}{3x_{J}^{2}}\right] \right.  \notag \\
&&+\left. \left( \widetilde{g}_{a}^{J}\right) ^{2}\left[ \frac{1}{3x_{J}}\left(
9\Delta _{J}+1\right) +\frac{1}{3x_{J}^{2}}+\frac{2}{x_{J}}\ln (x_{J})\right] \right\} .
\label{derivative-Z'-Z'-self}
\end{eqnarray}

Taking again the limit $M_{Z^{\prime },K}^{2}\rightarrow 0$ in order to obtain the leading $m_{J}$ term, $\Sigma _{Z^{\prime
}Z^{\prime }}^{\prime }$ in Eq. (\ref{derivative-Z'-Z'-self}) becomes

\begin{eqnarray}
\Sigma _{Z^{\prime }Z^{\prime }}^{\prime } &\approx &\left( \frac{g_{L}}{%
4\pi C_{W}}\right) ^{2}\left( \widetilde{g}_{a}^{J}\right) ^{2}\left[ 2\ln
(x_{J})\right] ,  \label{derive-leading}
\end{eqnarray}%

\vspace{0.3cm}

\begin{figure}[t]
\centering \includegraphics[scale=0.9]{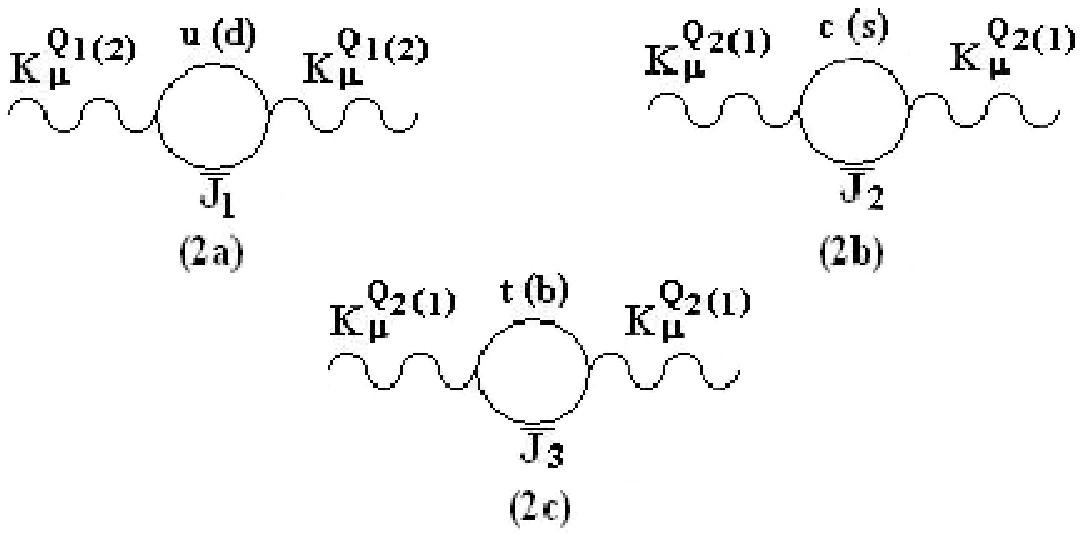}
\caption{\textsf{\ {\protect\small $K^{Q_{1,2}}$ gauge boson polarizations}}}
\label{fig-oblicua-charge}
\end{figure}

\begin{figure}[t]
\centering \includegraphics[scale=0.7]{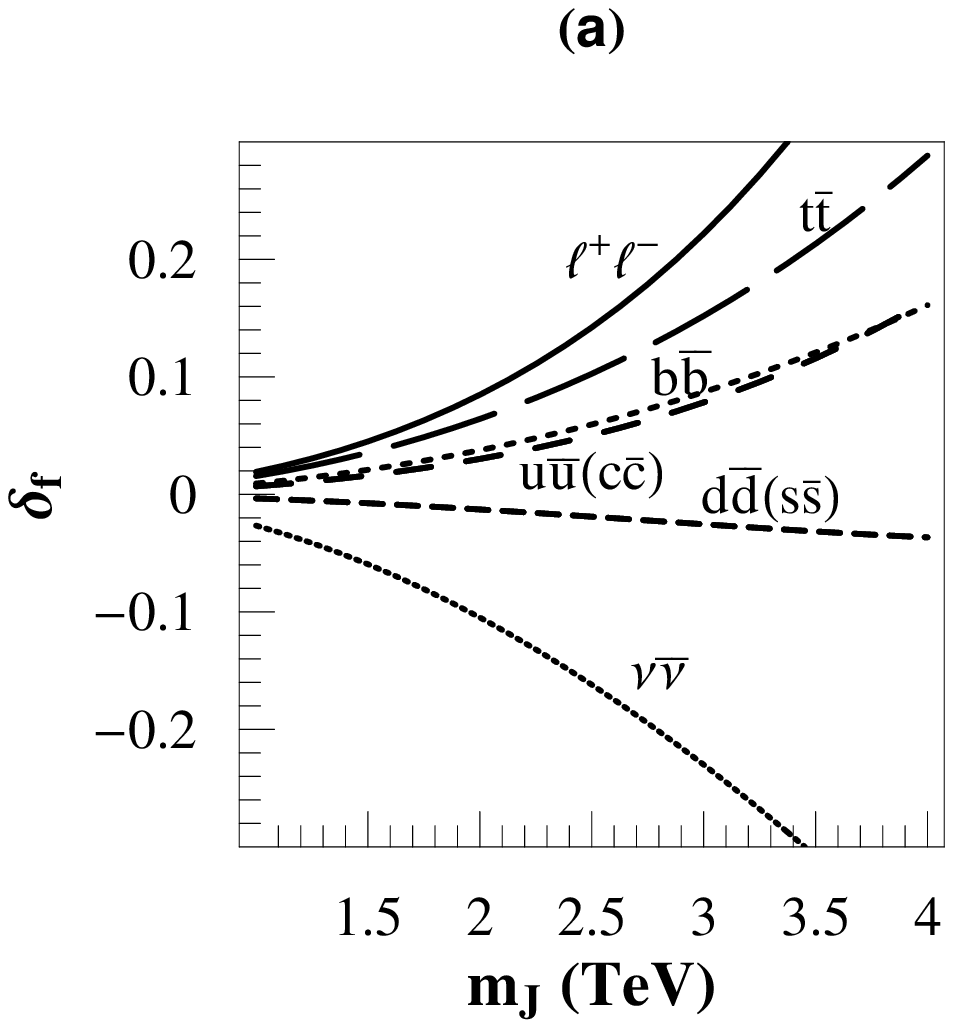} %
\includegraphics[scale=0.7]{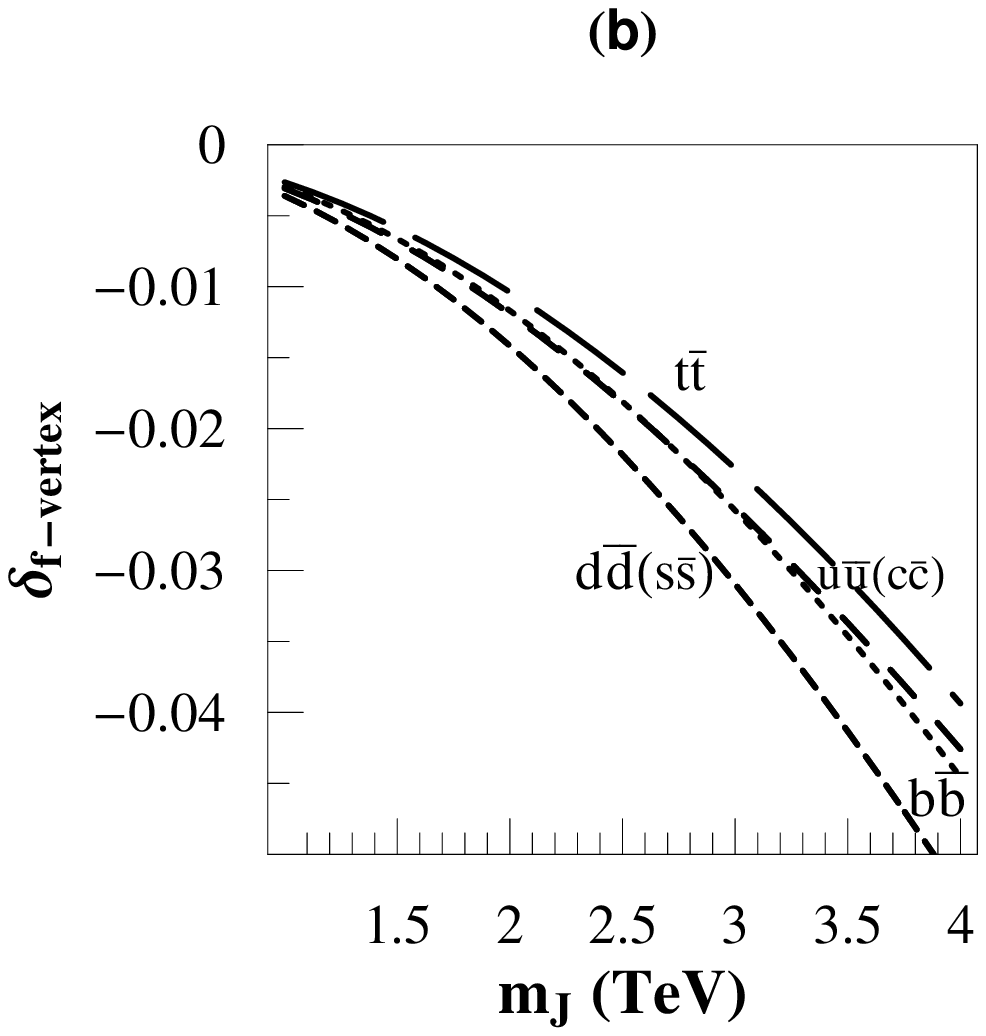}
\caption{\textsf{\ {\protect\small Deviation of the $Z^{\prime}$ decay width due to (a) the total radiative corrections as a function of the $J$ mass, and (b) the vertex corrections in the quark sector.}}}
\label{fig-branching-1}
\end{figure}

\begin{figure}[t]
\centering \includegraphics[scale=0.7]{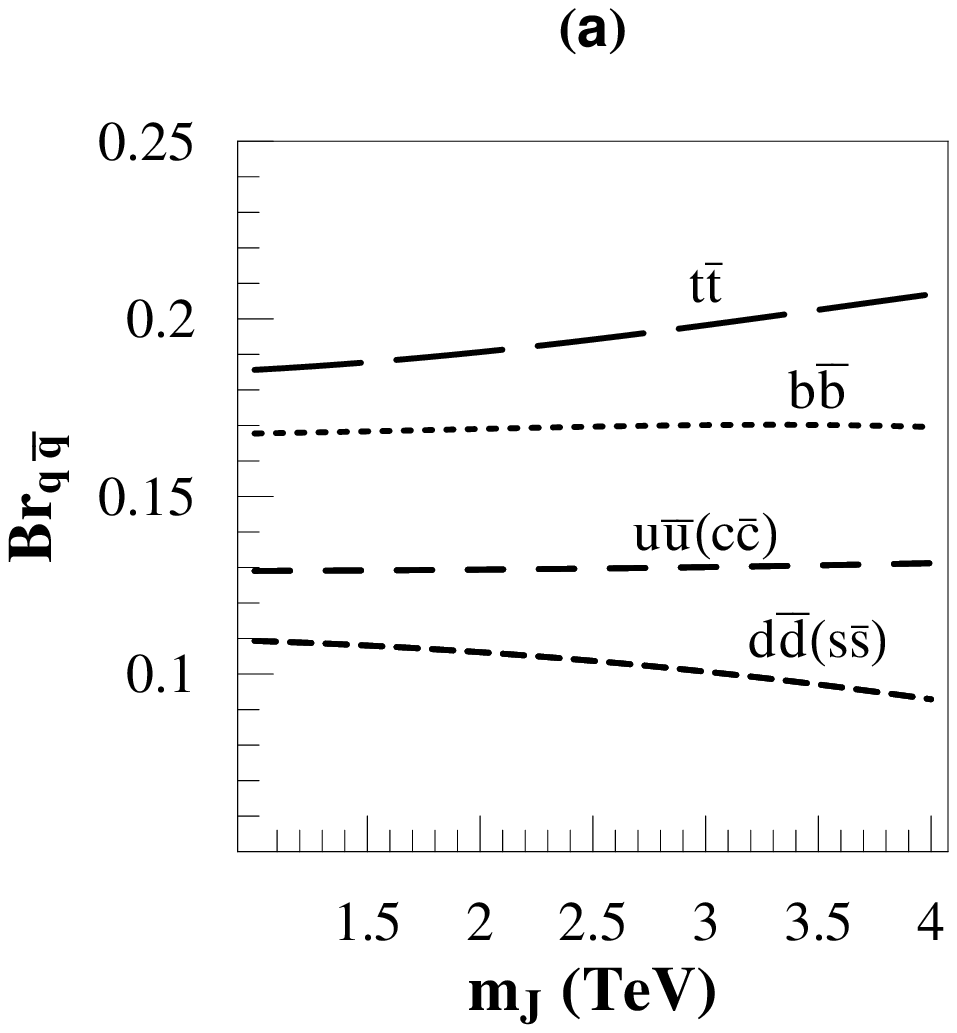} %
\includegraphics[scale=0.7]{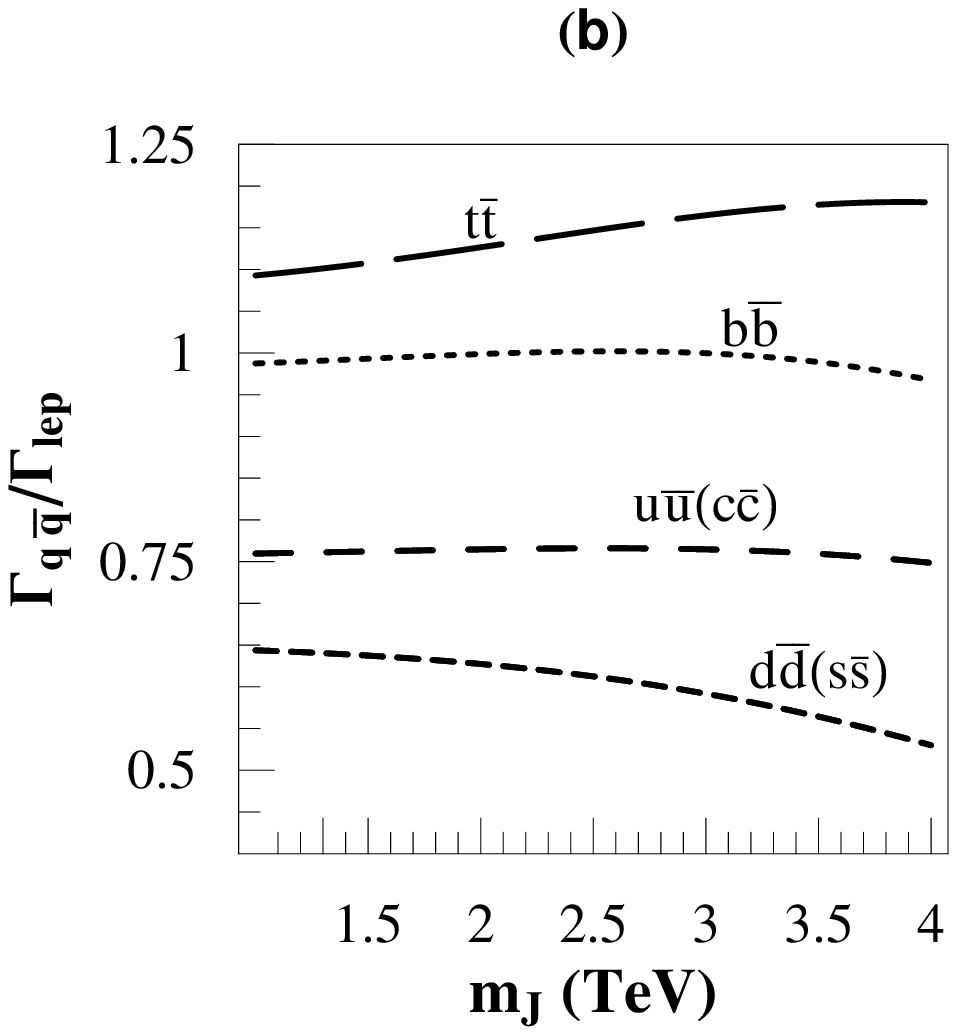}
\caption{\textsf{\ {\protect\small Branching ratios of $Z^{\prime} \rightarrow q\overline{q}$ (a) normalized to the width into all fermions and (b) normalized to the width into all leptons.}}}
\label{fig-branching-2}
\end{figure}

\begin{figure}[t]
\centering \includegraphics[scale=0.5]{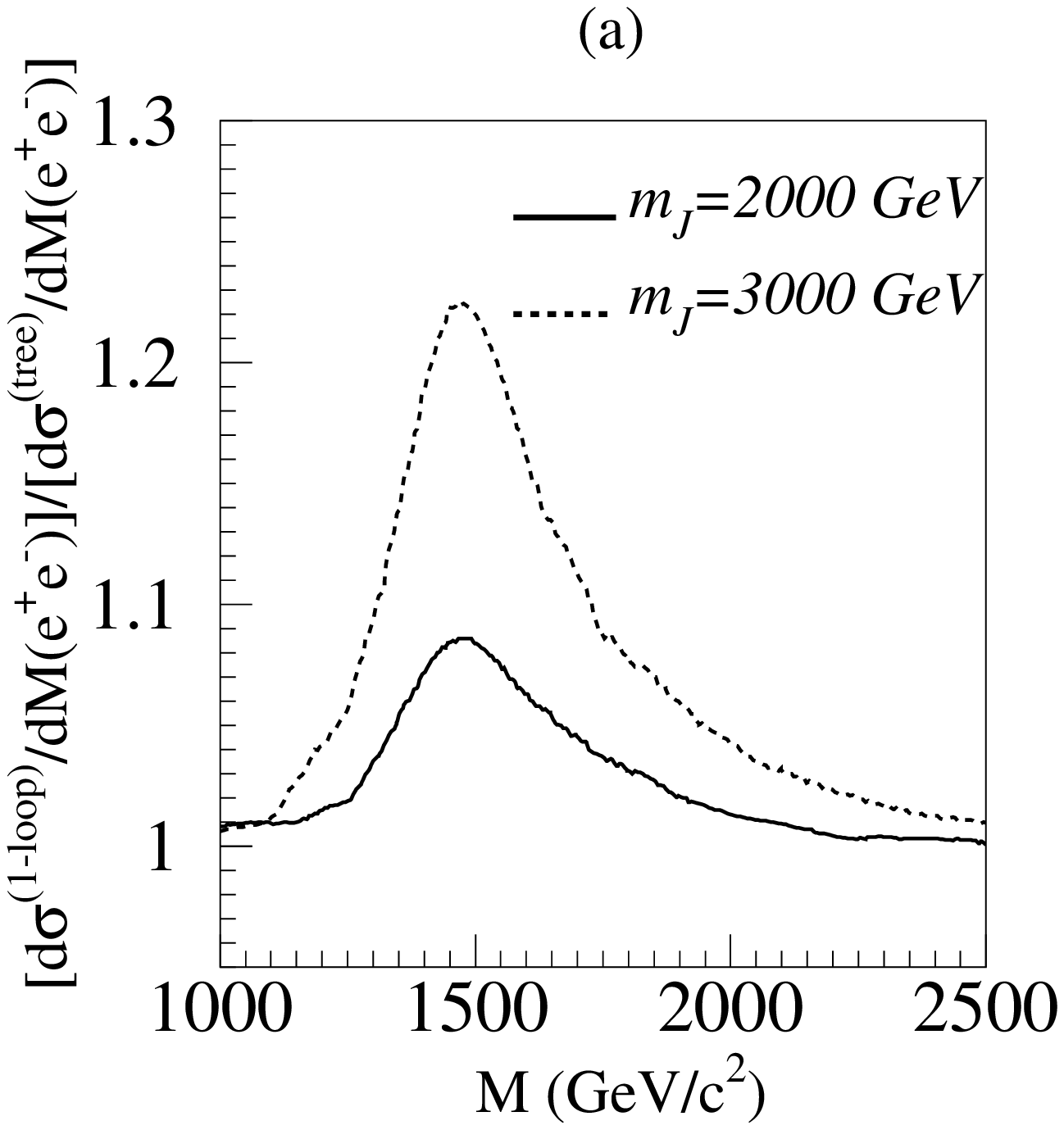} %
\includegraphics[scale=0.5]{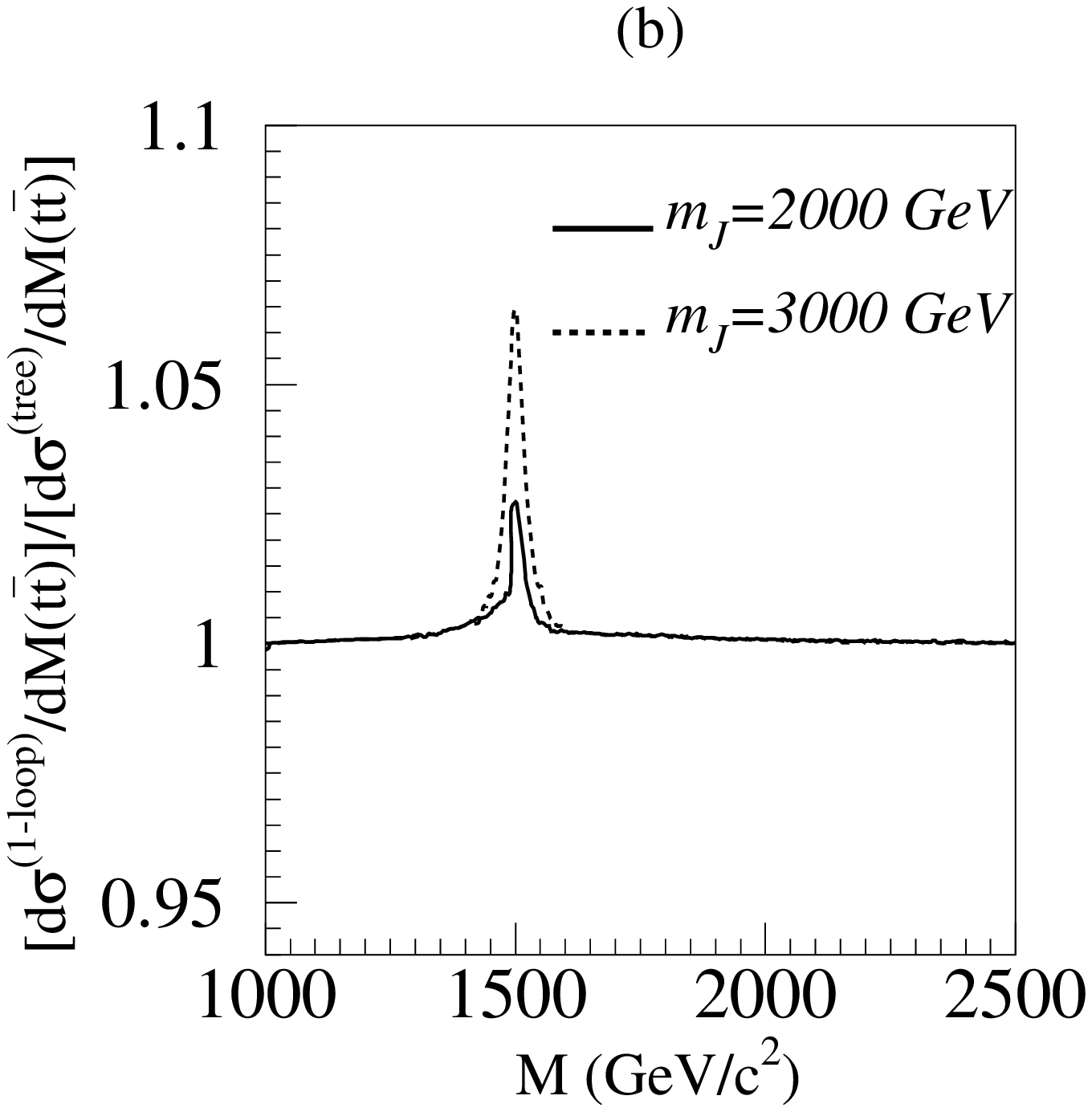}
\caption{\textsf{\ {\protect\small Ratio of the one-loop to tree-level invariant mass distribution for (a) di-electron final state and (b) top system for the CERN LHC collider for two values of $m_{J}$: $2$ TeV (continuous line) and $3$ TeV (dashed lines).}}}
\label{fig-distr}
\end{figure}

\ding{173} \underline{\textbf{The $K^{\pm Q_{1,2}}$ mass renormalization}}: 

The mass of the $K^{\pm Q_{1,2}}$ gauge bosons can be renormalized through the self-energy diagrams shown in Fig. \ref{fig-oblicua-charge}. Similar to the $Z^{\prime}$ case described before, the renormalized $K_{\mu}$ mass is defined as

\begin{equation}
M_{K0}^{2}=M_{K}^{2}\left( 1+3\Pi _{K}(M%
_{K}^{2})\right) .  \label{mass-K-renor}
\end{equation}

Since the $K$ gauge bosons couple to the $J$ quarks in the same way as the SM $W$ bosons couple to the $t$ quark, as seen in Eq. (\ref{L-Dirac-fisico}), the vacuum polarization of the $K$ bosons in the static limit is the same as the $W$ boson. Then, 

\begin{equation}
\Pi _{K}(0)=-\left( \frac{g_{L}}{4\pi }\right) ^{2}\frac{m_{J}^{2}}{M_{K}^{2}}\left[ \frac{%
3\Delta _{J}}{2}+\frac{3}{4}\right] .  \label{K-K-self}
\end{equation}

\vspace{0.3cm}

\ding{174} \underline{\textbf{The coupling constant renormalzation}}:

At tree-level, the coupling associated to the $W^{\pm }$ gauge bosons is characterized by the Fermi constant $G_{F}=g_{L}^{2}/4\sqrt{2}M_{W}^{2}$. Similarly, we can define a like-Fermi constant associated to the heavy $%
K^{\pm Q_{1,2}}$ gauge bosons, given by

\begin{equation}
G_{F}^{\prime }=g_{L}^{2}/4\sqrt{2}M_{K}^{2}, \label{K-fermi}
\end{equation}

with $M_{K}=M_{K^{Q_{1}}}=M_{K^{Q_{2}}}$ given by Eq. (\ref{charged-mases}). We can see that the SM Fermi constant and the $K-$Fermi constant are related by $G_{F}^{\prime }/G_{F}=M_{W}^{2}/M_{K}^{2}\ll 1$. Since the $K^{\pm Q_{1,2}}$ bosons decouple at low energy from the light matter, the $G_{F}^{\prime }$ constant is important only at the $M_{K}$ scale energy. This new Fermi constant is renormalized due to the $K$ vacuum polarization as

\begin{equation}
G_{F}^{\prime }=\frac{g_{L}^{2}}{4\sqrt{2}M_{K}^{2}}=\frac{g_{L0}^{2}}{4\sqrt{2}%
M_{K0}^{2}}\left( 1+3\Pi _{K}(0)\right) ,  \label{fermi'-renorm}
\end{equation}%

with $\Pi _{K}$ given by Ec. (\ref{K-K-self}) in the static limit. Using the expression from Eq. (\ref{scheme}) for the bare and renormalized quantities in (\ref{fermi'-renorm}), we obtain

\begin{equation}
\frac{g_{X0}^{2}}{T_{W0}^{2}}=\frac{g_{X}^{2}}{T_{W}^{2}}\left( 1+\Delta \rho _{j}\right), \label{coupling-renor}
\end{equation}

where we define the oblique parameter as

\begin{equation}
\Delta \rho _{j}=3\Pi _{Z^{\prime}}(0)-3\Pi _{K}(0)=\left( \frac{g_{X}}{4\pi T_{W}}\right)
^{2}\left[ 3x_{J}+2\ln (x_{J})%
\right], \label{oblique-parameter1}
\end{equation}

where we use the expressions for $\Pi _{Z^{\prime}}(0)$ and $\Pi _{K}(0)$ given by Eqs. (\ref{polar-leading}) and (\ref{K-K-self}), and the definition in Eq. (\ref{scheme}). We see that the divergent term $\Delta _J$ from the vacuum polarization is removed.

\vspace{0.3cm}

\ding{174} \underline{\textbf{The Weinberg angle renormalization}%
}: The Weinberg angle is defined by Eq. (\ref{weinberg}). In order to obtain the expression for the renormalized angle, we find a relation between the Weinberg angle and the renormalized gauge boson masses $M_{K}$ and $M_{Z^{\prime}}$. This can be done if we replace the relation $(g_{L}/g_{X})^2$ obtained from Eq. (\ref{weinberg}) in the Eq. (\ref{scheme}), from where we obtain the renormalized Weinberg angle

\begin{equation}
T_{W}^{2}=\frac{1}{\beta ^{2}}\left(1-\frac{4}{3}\frac{M_{K}^2}{M_{Z^{\prime}}^2} \right). \label{reweinberg}
\end{equation} 

Using the relations from Eqs. (\ref{mass-z'-renorm}) and (\ref{mass-K-renor}), and the oblique parameter definiton from Eq. (\ref{oblique-parameter1}), we obtain from Eq. (\ref{reweinberg}) the relation between the renormalized and the bare angle

\begin{equation}
T_{W0}^{2}=\overline{T_{W}^{2}}=T_{W}^{2}\kappa ^{\prime}_{f}=T_{W}^{2}\left(1+\epsilon \Delta \rho _{j} \right), \label{weinberg-renor}
\end{equation}

where

\begin{equation}
\epsilon = \frac{1-\beta ^{2}T_{W}^{2}}{\beta ^{2}T_{W}^{2}}. \label{epsilon}
\end{equation}
 
\vspace{0.3cm}

With the above contributions, the tree-level lagrangian for the $Z^{\prime}$ couplings in Eq. (\ref{L-Dirac-fisico}) becomes

\begin{eqnarray}
\mathcal{L}^{Z^{\prime}} &=&\overline{f}\gamma ^{\mu }\left[ \frac{g_{L0}}{2C_{W0}}%
\left( \widetilde{g}_{v0}^{f}-\widetilde{g}_{a0}^{f}\gamma _{5}\right) +(3%
\Pi _{Z^{\prime }Z})\frac{g_{L0}}{2C_{W0}}\left(
g_{v0}^{f}-g_{a0}^{f}\gamma _{5}\right) \right.   \notag \\
&&+\left. (3\Pi _{Z^{\prime }\gamma })g_{L0}S_{W0}Q_{f}\right]
fZ_{\mu }^{\prime }\mathcal{Z}_{3}^{-1/2},  \label{L-neutro-corregido}
\end{eqnarray}%

where the self-energy $\Sigma _{Z^{\prime }Z^{\prime }}$ from the diagram in Fig. (1a) is absorbed into the field renormalization constant $\mathcal{Z}_{3}$ defined in Eq. (\ref{mass-z'-renorm}), and which can be approximated at first order to $%
\mathcal{Z}_{3}^{-1/2}\approx 1-(1/2)3\Sigma _{Z^{\prime }Z^{\prime
}}^{\prime }$. The vector and axial-vector couplings $\widetilde{g}_{v,a0}^{f}$ are the same as Eqs. (\ref{Zprima-couplings}) but with the bare parameters, which are related to the renormalized quantities through Eqs. (\ref{coupling-renor}) and (\ref{weinberg-renor}). Since we are interested to obtain the leading $m_{J}^2$ contributions, the $\Pi _{Z^{\prime }Z}$ and $\Pi _{Z^{\prime }\gamma }$ polarizations are neglegible, as we disscused before. Then, we obtain the following lagrangian

\begin{equation}
\mathcal{L}^{Z^{\prime}}=\frac{g_{X0}}{4T_{W0}}\overline{f}\gamma ^{\mu }\left( 
\widetilde{g}_{v,eff}^{f}-\widetilde{g}_{a,eff}^{f}\gamma _{5}\right)
fZ_{\mu }^{\prime }\left( 1-\frac{3}{2}\Sigma
_{Z^{\prime }Z^{\prime }}^{\prime }\right) ,  \label{L-neutro-corregido-2}
\end{equation}%

where we have re-parametrized the vector and axial couplings as

\begin{eqnarray}
\widetilde{g}_{v,a,eff}^{U} &=&%
\mbox{\small$ \frac{1}{\sqrt{3}}\left[ diag\left( -1,1,1\right) +\frac{\beta
\overline{T_{W}^{2}}}{\sqrt{3}}\right] \pm 2Q_{U}\beta \overline{T_{W}^{2}} , $}  \notag
\\
\widetilde{g}_{v,a,eff}^{D} &=&%
\mbox{\small$\frac{1}{\sqrt{3}}\left[ diag\left( -1,1,1\right) +\frac{\beta
\overline{T_{W}^{2}}}{\sqrt{3}}\right] \pm 2Q_{D}\beta \overline{T_{W}^{2}} ,$} \notag
\\
\widetilde{g}_{v,a,eff}^{N} &=&%
\mbox{\small$-\frac{1}{\sqrt{3}}-\beta
\overline{T_{W}^{2}},$} \notag
\\
\widetilde{g}_{v,a,eff}^{L} &=&%
\mbox{\small$-\frac{1}{\sqrt{3}}-\beta
\overline{T_{W}^{2}}\pm 2Q_{L}\beta \overline{T_{W}^{2}} ,$}  \label{Zprima-couplings2}
\end{eqnarray}

with $\overline{T_{W}^{2}}$ given by Eq. (\ref{weinberg-renor}). Using Eqs. (\ref{derive-leading}) and (\ref{coupling-renor}), we finally obtain

\begin{equation}
\mathcal{L}^{Z^{\prime}}=\frac{g_{X}}{4T_{W}}\sqrt{\rho _{f}^{\prime}}\overline{f}\gamma ^{\mu }\left( 
\widetilde{g}_{v,eff}^{f}-\widetilde{g}_{a,eff}^{f}\gamma _{5}\right)
fZ_{\mu }^{\prime },  \label{L-neutro-corregido-3}
\end{equation}%

with $\rho _{f}^{\prime}=1+\Delta \rho _{j}$, where for $\Delta \rho _{j}$ we have taken only the $m_{J}^2$ term of Eq. (\ref{oblique-parameter1}), i.e.

\begin{equation}
\Delta \rho _{j}\approx 3\left( \frac{g_{X}}{4\pi T_{W}}\right)
^{2}x_{J}. \label{oblique-parameter2}
\end{equation}

\subsection{Vertex corrections\label{appendixA2}}

\begin{figure}[t]
\centering \includegraphics[scale=0.9]{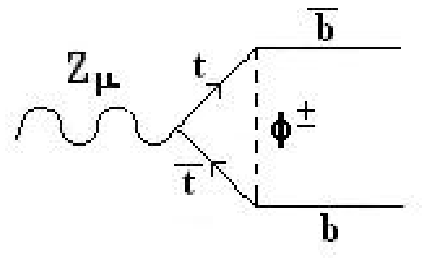}
\caption{\textsf{\ {\protect\small Goldstone-exchange contribution to the $Zb\overline{b}$ vertex.}}}
\label{fig:goldstone-SM}
\end{figure}

Now, we will consider the diagrams from Figs. (1d,e,f). In the SM, the vertex corrections are considered because of the top quark effects that arise from $W-$exchange in the one-loop correction to the $Zb\overline{b}$ vertex. These corrections in the renormalizable 't Hooft-Feynman gauge are depicted by a set of diagrams which contains $W^{\pm}$ gauge bosons and charged Goldstone $\phi ^{\pm}$ which are absorbed by $W$ mass terms after the spontaneous symmetry breaking. The explicit calculation of these contributions in the SM is obtained in for example Ref. \cite{bernabeu-91}. However, the dominant $m_{t}^2$ contribution comes entirely from one diagram: the corresponding to the exchange of a virtual $\phi ^{\pm}$ shown in Fig. \ref{fig:goldstone-SM} \cite{bernabeu-88}.

\begin{figure}[t]
\centering \includegraphics[scale=0.9]{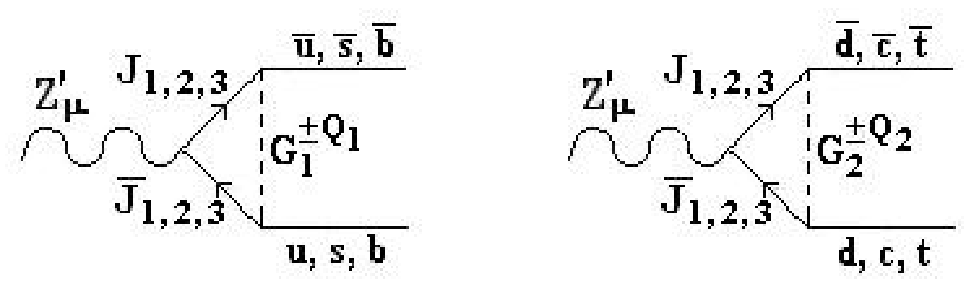}
\caption{\textsf{\ {\protect\small Goldstone-exchange contribution to the $Z^{\prime}q\overline{q}$ vertex.}}}
\label{fig:goldstone-331}
\end{figure}

On the other hand, the $Z^{\prime}$ and $K^{Q_{1,2}}$ couplings in Eq. (\ref{L-Dirac-fisico}) are the same as the $Z$ and $W^{\pm}$ couplings but with the change of $g_{v,a}$ by $\widetilde{g}_{v,a}$. Thus, the integrals from the loop corrections will have the same behaviour as the SM calculation but constant factors that depend on the parametrization of $\widetilde{g}_{v,a}$. From the above discussion, we can conclude that the hard mass term $m_{J}^2$ is only contained in the diagrams from Fig. \ref{fig:goldstone-331}, where $G_{1,2}^{Q_{1,2}}$ are the Goldstone bosons associated to the $K^{Q_{1,2}}$ gauge bosons, respectively, as described in Ref. \cite{DMO-05} (where $G_{1,2}^{Q_{1,2}}$ are marked as $\phi _{2,3}^{Q_{1,2}}$). The Yukawa interaction associated to the Goldstone bosons is

\begin{eqnarray}
\mathcal{L}_Y&=&\frac{g_{L}m_{J}}{\sqrt{2} M_{K}}\left[\left(\overline{u_{1L}},\overline{J_{2R}},\overline{J_{3R}}\right)G_{1}^{Q_{1}} \left(\begin{array}{c}J_{1R} \\ d_{2L} \\ d_{3L} \end{array} \right) \right. \notag \\
&&+\left. \left(\overline{d_{1L}},\overline{J_{2R}},\overline{J_{3R}}\right)G_{2}^{Q_{2}} \left(\begin{array}{c}J_{1R} \\ u_{2L} \\ u_{3L} \end{array} \right)+ h.c \right].
\end{eqnarray}   

The calculation of diagrams in Fig. \ref{fig:goldstone-331} yield contributions proportional to $x$, $x\ln (x)$, $\ln (x)$ and divergent terms $\Delta$, where $x=(m_{J}/M_{K})^{2}$. However, if we calculate all the one-loop vertex diagrams, as done in Refs. \cite{bernabeu-91} and \cite{bernabeu-88} in the case of SM, the $x\ln (x)$ and $\Delta$ terms cancels out. Thus, if we isolate only the $x$ terms, we get the following value for the vertex correction of any of the diagrams in Fig. \ref{fig:goldstone-331}

\begin{equation}
\frac{ig_{L}}{2C_W}\Gamma _{\mu}^{(1)}= \frac{g_{L}^{2}}{32\pi ^{2}}F_{q}(m_{J_{1,2,3}})\left[ 1-\gamma_{5}\right], \label{vertex-function}
\end{equation}

where $F_{q}(m_{J_{1,2,3}})\approx (m_{J}/M_{K})^{2}(g_{X}C_{W}/\sqrt{3}g_{L}T_{W})diag(1,-1,-1)$. Thus, using Eq. (\ref{vertex-function}) we get the following counterterm 

\begin{equation}
\delta \mathcal{L}_{J-vertex}=\left( \frac{g_{L}}{2C_W}\right) \left( \frac{g_{L}^{2}}{32\pi ^{2}}\right)F_{q}(m_{J_{1,2,3}})\overline{q} \gamma ^{\mu } \left[ 1-\gamma_{5}\right] qZ^{\prime}_{\mu}.
\end{equation}

Adding the above contribution to the lagrangian (\ref{L-neutro-corregido-3}), and taking into account that $(m_{J}/M_{K})^{2}=(4g_{X}^{2}/3g_{L}^{2}T_{W}^{2})x_{J}$ and the definition from Eq. (\ref{oblique-parameter2}), we obtain the total lagrangian for quarks

\begin{equation}
\mathcal{L}^{Z^{\prime}}=\frac{g_{X}}{4T_{W}}\overline{q}\gamma ^{\mu }\left[ 
\sqrt{\rho _{q}^{\prime}}\widetilde{g}_{v,eff}^{q}+\delta \widetilde{g}_{v}^{q}-\left( \sqrt{\rho _{q}^{\prime}}\widetilde{g}_{a,eff}^{q}+\delta \widetilde{g}_{a}^{q}\right) \gamma _{5}\right]
qZ_{\mu }^{\prime },  \label{L-neutro-corregido-4}
\end{equation}

where the vertex correction is absorbed into the vector and axial-vector coupling correction $\delta \widetilde{g}_{v,a}^{q}=(4/9)\Delta \rho _{j}(1/\sqrt{3})diag(1,-1,-1)$. The lagrangian (\ref{L-neutro-corregido-4}) can be written as

\begin{equation}
\mathcal{L}^{Z^{\prime}}=\frac{g_{X}}{4T_{W}}\sqrt{\overline{\rho _{q}^{\prime}}}\overline{q}\gamma ^{\mu }\left( 
\overline{\widetilde{g}_{v,eff}^{q}}-\overline{\widetilde{g}_{a,eff}^{q}}\gamma _{5}\right)
qZ_{\mu }^{\prime },  \label{L-neutro-corregido-5}
\end{equation}%

with $\overline{\rho _{q}^{\prime}}=1+\Delta \rho _{j}-(8/9)\Delta \rho _{j}$ and $\overline{\widetilde{g}_{v,eff}^{q}}$ the same as $\widetilde{g}_{v,eff}^{q}$ in Eq. (\ref{Zprima-couplings2}) but with $\kappa ^{\prime}$ changed by $\overline{\kappa ^{\prime}_{q}}=1+\epsilon \Delta \rho _{j}+(4/9)\Delta \rho _{j}$.

%Figure Captions

\end{document}